\documentclass[sn-basic]{sn-jnl}

\usepackage{amsmath}
\usepackage{amssymb}
\usepackage{amsfonts}
\usepackage{amsthm}
\usepackage{mathtools}
\usepackage{enumitem}
\usepackage{bm}
\usepackage{rotating}
\usepackage{booktabs}
\usepackage{array}
\usepackage{tabularx}
\usepackage{makecell}
\usepackage{multirow}
\usepackage{placeins}

\newcolumntype{Y}{>{\raggedright\arraybackslash}X}
\newcommand{\mat}[1]{\begin{pmatrix} #1 \end{pmatrix}}

\newcommand{\interest}{\psi}

\newcommand{\ttheta}{\bm{\theta}}
\newcommand{\pphi}{\bm{\phi}}

\newcommand{\llambda}{\bm{\lambda}}

\newcommand{\pivot}{m_\interest} 

\newcommand{\xx}{\mathbf{x}}
\newcommand{\uu}{\mathbf{\ttheta}}
\newcommand{\kk}{\mathbf{\bm{\nu}}}
\newcommand{\zz}{\mathbf{z}}

\newcommand{\XX}{\mathbf{X}}
\newcommand{\ZZ}{\mathbf{Z}}

\newcommand{\uuu}{\mathbf{u}}
\newcommand{\UU}{\mathbf{U}}

\newcommand{\vv}{\mathbf{v}}
\newcommand{\VV}{\mathbf{V}}

\newcommand{\mboundary}{\widetilde{m_z}}
\newcommand{\mboundaryNoU}{\overline{m}_z}
\newcommand{\mswap}{q_z}

\newcommand{\uuhat}{\hat{\uu}}

\newcommand{\zpivot}{z_\mathrm{p}}
\newcommand{\znuis}{\zz_\mathrm{n}}
\newcommand{\Zpivot}{Z_\mathrm{p}}
\newcommand{\Znuis}{\ZZ_\mathrm{n}}
\newcommand{\cinet}{\interest_{\mathrm{CI}}}
\newcommand{\wi}[1]{\phi^{(#1)}}
\newcommand{\NN}{f_{\bm{\phi}}}
\newcommand{\NNi}[1]{f^{(#1)}_{\wi{#1}}}
\newcommand{\RR}{\mathbb{R}}
\newcommand{\Normal}{\mathcal{N}}
\newcommand{\Loss}{\mathcal{L}}
\newcommand{\KL}{D_{\mathrm{KL}}}
\newcommand{\E}{\mathbb{E}}

\newcommand{\pz}{p_\zz(\zz)}
\newcommand{\px}{p_\xx(\xx)}
\newcommand{\qz}{q_\zz(\zz)}
\newcommand{\qx}{q_\xx(\xx)}
\newcommand{\qbar}{\bar{q}_\xx}
\newcommand{\qbarx}{\qbar(\xx)}

\newcommand{\Astar}{A^\ast}

\newcommand{\sx}{g_\xx}
\newcommand{\s}{\sx}

\newcommand{\nx}{n_\xx}
\newcommand{\nknown}{n_\kk}
\newcommand{\nunknown}{n_\uu}
\newcommand{\nG}{n_G}

\newcommand{\paramset}{\Theta}
\newcommand{\innerTheta}{\paramset_{\mathrm{inner}}}
\newcommand{\outerTheta}{\paramset_{\mathrm{outer}}}
\newcommand{\xset}{\mathcal{X}}
\newcommand{\xtarget}{\xset_{\mathrm{target}}}
\newcommand{\boundingbox}{\mathcal{B}}

\newcommand{\Cov}{\mathrm{Cov}}

\newcommand{\dd}{\mathrm{d}}

\newcommand{\score}{s}
\newcommand{\info}{I}
\newcommand{\sscore}{\mathbf{\score}}
\newcommand{\iinfo}{\mathbf{\info}}
\newcommand{\effscore}{\score_{\interest \cdot \llambda}}
\newcommand{\effinfo}{\info_{\interest \cdot \llambda}}

\newcommand{\interesthat}{\hat{\interest}}
\newcommand{\lambdahat}{\hat{\llambda}}
\newcommand{\pivotlocal}{\pivot^{\textrm{local}}}
\newcommand{\ellhat}{\hat{\ell}}

\begin{document}
	
	\title[Pivoting with normalizing flows]{Likelihood-free inference with nuisance parameters through normalizing flows}
	
	\author*[1]{\fnm{Phil} \sur{Assheton}} \email{phil@statsadvice.com}
	
	\affil*[1]{\orgname{statsadvice.com}, \city{Berlin}, \country{Germany}}
		
	\abstract{We present a simple decomposition of a neural-network-based normalizing flow that naturally uncovers a pivotal statistic (or something close) in the presence of nuisance parameters, based only on a sample generator from the distribution of interest.  We show that the statistic is near-pivotal in the sense of minimum average KL-divergence of its $p$-values versus uniform and we argue that it can be expected to have good power when the dimension of the statistic equals the dimension of the parameter.  It is able to incorporate prior knowledge about group invariances such as translation and scale.  It can discover the one-sample $t$-test almost exactly, outperforms the Welch test in terms of worst-case size over a constrained variance-ratio range and achieves good calibration on partial biserial correlations, while showing higher power (and being much faster) on small-to-moderate samples than profile likelihood-ratio techniques.}
	
	\keywords{pivotal statistic, nuisance parameters, likelihood-free inference, simulator-based inference}
	
	\maketitle
	
	\section{Introduction}
	\label{section intro}
	
	\subsection{Context}
	\label{section lit review}
	
	Frequentist inference in the presence of nuisance parameters ideally seeks a ``similar test'': one whose size is invariant across all possible configurations of the nuisance parameters.  In many \textit{specific} cases, clever tricks have been found to compute pivotal statistics from the sample, whose distribution does not vary with the nuisance parameter(s).  A well-known example of this is the $t$-test, where the goal is to perform inference on the mean $\mu$ of a normal distribution, regardless of its variance, $\sigma^2$; a $t$-statistic is computed, which effectively cancels the effect of $\sigma$ out, leaving a statistic that is pivotal and a test that is similar across all values of $\sigma$.
	
	On the other hand, a number of more ``general'' approaches do exist, whose goal is to be able to perform inference across a broad range of parametric parent distributions.  Owing to the complexity of the problem of finding a similar test across an entire space of nuisance parameters, these each rely on different constraints or approximations.
	
	The profile likelihood ratio (LR) is the basis for many of these.  The profile LR compares the likelihood at the maximum likelihood estimate (MLE) to that at a null-constrained MLE, where the interest parameter is constrained to equal some null value.  In the large sample limit, it can be shown that $-2\log LR$ converges in distribution to a $\chi^2_1$ distribution \citep{wilks1938large}.  That is, its limiting distribution is, to first order, independent of nuisance-parameter values.
	
	An alternative general approach is the parametric bootstrap \citep{davison1997bootstrap}, in which many samples are drawn from a fitted model, and used to approximate the sampling distribution.  One may, for example, draw many datasets from a model using the MLE as its parameters, and map each dataset to a given scalar statistic; the spread of these bootstrap statistics can be used to estimate the standard error of that statistic.  Alternatively, sampling from a model with parameter set to the \textit{null-constrained} MLE can be used to generate a $p$-value against a given null hypothesis.  
	
	One popular statistic to bootstrap is the profile LR, since it is known to be highly informative about the interest parameter and also to have a limiting distribution that is pivotal.  This effectively combines the bootstrap and LR approaches: using the powerful and asymptotically-pivotal LR, without relying on the large-sample assumption required for the $\chi^2$ approximation.
	
	Nevertheless, the bootstrap is still only an approximation: the bootstrap distribution is drawn from the MLE, rather than the true population parameter.  Higher order refinements and iterating bootstraps-within-bootstraps can improve accuracy in some settings, but at the cost of substantially increased computation \citep{hall1988bootstrap}.
		
	In recent years, there has been growing interest in applying machine learning techniques to this problem.  A deep neural network (DNN) is an extremely flexible, differentiable function approximator that can be quite efficiently trained to optimize some loss function.  For example, \cite{coccaro2020dnnlikelihood} use DNNs as surrogates for complex likelihood functions, so that they may be used in LR or Bayesian approaches, among other possibilities.
	
	\cite{cranmer2016approximating} show that a neural net trained by cross-entropy to distinguish between samples generated under two different parameter values encodes (up to a transformation) the LR statistic, without needing to model the likelihood directly.  Instead it is trained on samples drawn from the model of interest; in many real-world cases, such a simulator is available where an explicit likelihood function is not.  In the presence of nuisance parameters, the same classifier-based ratio idea can be embedded in a profile-LR construction.   \cite{heinrich2022learning} trains a model to distinguish data from any given null vs data from any of a mixture of alternatives, all at an equal Fisher-metric distance from the given null.  He notes that this should target the test statistic with best average power (for this mixture) and show that it recovers the profile LR in cases where that is known to be most powerful.  
	
	For the final step from statistic to $p$-value, \cite{dalmasso2024likelihood} present a framework in which learned test statistics (such as their classifier-based LR-approximator called ACORE) are mapped to critical values (for confidence intervals/regions) by quantile regression and $p$-values by a probabilistic binary classifier.  They refer to this framework as ``Likelihood-free Frequentist Inference'' (LF2I), again stressing the broader range of applications where no likelihood function is required; only a simulator.  \cite{al2024amortized} take this a step further to model the entire curve of critical values by training a binary classifier to model the CDF of the test statistic given the parameter.  More specifically, they train the net by least squares to model the expectation of a binary indicator on whether samples are greater or less than a given threshold.
	
	Particularly relevant to our work, however, is that of \cite{louppe2017learning}. They, like us, attempt to train directly a neural net that will, by its design, identify a pivotal statistic. They do this by adapting the concept of generative adversarial networks (GANs) from \cite{goodfellow2014generative}. A secondary ``adversary'' net attempts to predict, by maximum likelihood, the values of the nuisance parameters from the output of the pivot-learning ``classifier'' net; the latter, in turn, aims to minimize the same predictive log likelihood, while simultaneously maximizing the log likelihood of a target label given the observed data. In the context of frequentist statistical inference, this target label could encode the hypothesis or parameter value of interest. By optimizing both objectives, the net uncovers a statistic that is pivotal with respect to the nuisance parameters while remaining optimally predictive of the target label.

	Training a neural network can be computationally expensive.  However, in many cases, once fit, the same network can be reused over and over again very cheaply, effectively paying off the initial investment.  For this reason, such approaches are often referred to as ``amortized'' \citep{zammit2025neural}.
	
	\subsection{Introducing NeuralCIs}
	\label{section intro neuralcis}
	
	In the following, we will introduce the first phase of a project called NeuralCIs, whose ultimate goal will be to use neural models to generate confidence intervals and regions.  In this first phase, we focus on the generation of $p$-values in the presence of nuisance parameters, but these are expected to be quite readily converted into confidence intervals by a further training step (see Section \ref{section confidence intervals}) and we hope will be generalizable to simultaneous confidence intervals and confidence regions over vector-valued interest parameter.  The code for the project can be followed on GitHub at \href{https://github.com/philassheton/neuralcis}{github.com/philassheton/neuralcis}.
	
	The idea is to train a single neural net that can solve an entire class of problems; once trained (this currently takes two to four hours on a modest RTX 3060 laptop GPU), the net can very quickly convert any dataset into a $p$-value (and ultimately confidence interval) against any of a range of values for the interest parameter; that is, the initial training cost is amortized.  
	
	Using a neural model in this way feels quite natural.  To generate a single $p$-value one must anyway effectively compare the one dataset of interest to all datasets that are probable under one null hypothesis; to generate a confidence interval extends this search out to many ``null hypotheses''.  Whereas bootstrapping would start afresh for each new dataset, fitting a single neural net to model all possible outcomes at all of a wide range of parameters means that that broad ``search'' across many samples from many parameter values can be shared and reused each time a new dataset arrives that fits a given problem; that initial cost is amortized.
	
	Furthermore, neural networks are quite efficiently differentiable, which makes them great for optimisation; we will take advantage of this in Section \ref{section confidence intervals} to propose a method to extend our $p$-values quite cleanly to confidence intervals.
	
	Our work adapts the idea of the normalizing flow (\cite{papamakarios2021normalizing}), which we will summarize in Section \ref{section normalizing flows}.  The key novel contribution is a decomposition of the normalizing flow that automatically yields an on-average near-pivotal statistic (Section \ref{section decomposing}).  We show that this statistic is near-pivotal in the sense that $p$-values derived from it have minimal average KL-divergence to uniform.  We further argue that it can be expected to be reasonably powerful by an implicit bias argument that becomes more concrete for transformation models at one extreme and in the large sample local asymptotic normal (LAN) limit at the other.  We assess both the calibration and power through simulation.
	
	In Section \ref{section normalizing flows}, we introduce the normalizing flow, how we implement it in NeuralCIs (somewhat different than normal), and how we can flip its input and output roles to provide amortized sampling from an unnormalized target density.  In Section \ref{section decomposing}, we detail the core decomposition that we will use to identify near-pivotal quantities, and how that can be used to generate $p$-values.  In Section \ref{section proof}, we explore the formal statistical properties of this decomposition, in particular, its expected false-positive rate behaviour as well as an implicit bias argument that it might be expected in many practically useful contexts to achieve good true-positive rates.  In Section \ref{section further considerations}, we outline further technical points central to our architecture (invariances, adding known values such as sample size, parameter sampling and neural network design).  In Section \ref{section experiments}, we present experimental methodology on one toy and three classical statistical problems, whose results we present in Section \ref{section results} and discuss in Section \ref{section discussion}.  A final conclusion is in Section \ref{section conclusion}.
	
	\section{Normalizing flows}
	\label{section normalizing flows}

	Normalizing flows are highly flexible probability density models that can approximate almost any continuous density encountered in practice on $\mathbb R^d$.  The idea is very simple: use a neural network to model an invertible change of variables, such that the distribution in the new variables is a simple one, commonly a standard normal.
	
	\subsection{Core concept}
	\label{section normalizing flows core concept}
	
	We will abbreviate $\Normal(\zz) \coloneqq \Normal(\zz; \mathbf 0, \mathbf I)$ and write a neural network mapping $\xx \mapsto \zz$ as $\zz = \NN(\xx)$.  Denote the true distribution of the inputs $\xx$ as $\px$ and of the net outputs $\zz$ as $\pz$.  We optimize the weights $\pphi$ of the network to bend samples from the true distribution $\pz$ as close as possible in distribution to the theoretical target distribution $\qz \coloneqq \Normal(\zz)$.  This induces a theoretical approximate distribution $\qx$ on $\xx$:
	\begin{align*}
		q_\xx(\xx) = q_\zz(\zz) \left| \mathrm{det} \frac{\partial \zz}{\partial \xx} \right| = \Normal \left( \NN(\xx) \right)  \left| \mathrm{det} \frac{\partial \NN(\xx)}{\partial \xx} \right|,
	\end{align*}
	whose fit we can optimize by minimizing the sum of the negative log likelihood across a large number of samples $\xx \sim \px$:
	\begin{align}
	\label{equation loss}
	-\log q_\xx(\xx)
	= -\log \left( q_\zz(\zz) \left| \mathrm{det} \frac{\partial \zz}{\partial \xx} \right| \right) 
	= -\log \Normal \left( \NN(\xx) \right) - \log \left| \mathrm{det} \frac{\partial \NN(\xx)}{\partial \xx} \right|.
	\end{align}
	
	This maximization of $\qx$ across samples from $\px$ is in the large-sample limit equivalent to minimizing the Kullback-Leibler (KL) divergence $\KL(p_\xx \| q_\xx) \equiv \KL(p_\zz \| \Normal)$ \citep{papamakarios2021normalizing}.
	
	\subsection{Invertibility and tractability}
	\label{section invertibility and tractability}
	
	In order to have tractable Jacobian determinants, a typical normalizing flow constrains each network layer to have a trivial Jacobian determinant (for example, by designing each layer to have a triangular Jacobian matrix), and to be invertible by construction.
	
	We instead compute Jacobians by back-propagation (chain-rule backward through the net) and enforce approximate invertibility by punishing zero and negative Jacobian determinant.  The latter is achieved by replacing the absolute value $|j|$ of the Jacobian determinant $j$ in (\ref{equation loss}) with $\max(j, \epsilon \sigma(j))$, with $\sigma(j) \coloneqq \tfrac{1}{1 + e^{-j}}$ and $\epsilon = 10^{-10}$ chosen to be small, without being small enough to erase the gradients of $\sigma$ below numerical precision.
	
	This approach is quite computationally expensive ($\mathcal O(\nx^3)$ in the number of statistics $\nx$) but our flows still train in a few hours on a very modest laptop GPU (RTX 3060 mobile).  We tolerate this extra cost because it affords us incredible flexibility in terms of introducing conditioning variables (Section \ref{conditional flows subsection}) and decomposing our flow (Section \ref{section decomposing}), without the need to experiment with exotic constrained architectures.  The key here is that the traditional normalizing flow (with those architectural constraints described above) is designed to work well at very high dimensions.  Ours is instead (at present) designed for high flexibility with relatively low-dimensional problems.  
	
	We may in later iterations of the project experiment with higher dimensional problems.  Apart from such architectural constraints, one particularly promising approach is offered in recent work by \cite{draxler2024free}.  They propose an approximation method for fitting free-form flows that is instead $\mathcal O(\nx)$ per data-point.  Furthermore, they are able to target a \textit{global} diffeomorphism via an extra inverse flow whose role is to reconstruct the input.  (Our approach can only at present target a \textit{local} diffeomorphism, which does leave scope for dysfunctional flows to be fitted, for example, by wrapping coordinates around a circular path back onto themselves.)
	
	\subsection{Conditional Normalizing Flows}
	\label{conditional flows subsection}
	
	With this more na\"ive approach, we can very easily make our normalizing flow model a conditional probability density $p(\xx|\uu)$, by simply adding $\uu$ as extra inputs to the network.  This allows it to fit a different mapping at each value of $\uu$, each representing that particular conditional ``slice'' of distribution.  Training proceeds as before, but with the extra $\uu$ inputs playing no role in the Jacobian $\tfrac{\partial \zz}{\partial \xx}$ computed in the loss function (\ref{equation loss}).  
	
	\subsection{``Denormalizing flows'' for amortized sampling}
	\label{section importance sampling flows}
	
	During training, we will need to generate a huge number of $\uu$ samples, which we hope to draw approximately according to a Jeffreys' prior, $p(\uu) \propto \sqrt{\det(I(\uu))}$ (further details in Section \ref{section param sampling}).  We can invert the logic of a traditional normalizing flow to train a model that, once fitted, can rapidly draw an unlimited supply of samples from such a differentiable unnormalized target density \citep{rezende2015variational, papamakarios2021normalizing}.
	
	We will again use $p_\zz$ and $p_\xx$ to refer to the true distribution of our $\zz$ and $\xx$ \textit{samples} and $q_\zz$, $q_\xx$ to mean some theoretical target distribution that we want to bend them toward.  The key intuition here is that we reverse all the roles, compared with the core normalizing flow concept described in Section \ref{section normalizing flows core concept}.  The samples we feed into the net are standard normal $\zz \sim p_\zz \coloneqq \Normal(0, 1)$, and the outputs will aim to be distributed according to some distribution $\xx \sim q_\xx \coloneqq c\qbar$, where we only know $\qx$ up to some multiplicative constant $c$.  We aim then to learn the mapping $\zz \mapsto \xx$.
	
	We rewrite (\ref{equation loss}) as
	\begin{align*}
		-\log \qz 
		= -\log \left(c\qbarx \left|\det\frac{\partial \xx}{\partial \zz} \right|\right)
		= -\log \qbar(\NN(\zz)) - \log \left|\det \frac{\partial \xx}{\partial \zz} \right| - \log c,
	\end{align*}
	and we may discard the constant $\log c$ term without affecting the optimal value for $\phi$.
	
	As with the regular normalizing flow, maximizing $\qz$ across a large enough sample of $\zz \sim p(\zz)$ is equivalent to minimizing $\KL(p_\zz, q_\zz) \equiv \KL(p_\xx, q_\xx)$.  Intuitively, the determinant encourages the outputs $\xx$ to spread out and explore the region of support of $q_\xx$, while the $\qbar$ component encourages them to cluster wherever $\qbar$ is high.
	
	In the literature, this is still viewed as a normalizing flow, but \textit{fit} by minimizing \textit{reverse} KL divergence.  As a shorthand in this paper, we will refer to this architecture as a ``\textbf{denormalizing flow}'', to signify that it maps samples \textbf{from} a normal distribution to the target, rather than vice versa.  This name distinguishes it from the form described in Section \ref{section normalizing flows core concept}, which forms the basis of most of the work in this paper; the ``denormalizing flow'' is used only for parameter sampling during training.
		
	\section{Decomposing the normalizing flow}
	\label{section decomposing}
	
	The key novel contribution of this paper is a very simple decomposition of the normalizing flow that naturally uncovers pivotal quantities and their distributions.
		
	Suppose that the parameter $\uu$ has dimension $\nunknown$ with scalar interest parameter $\interest(\uu)$; all other dimensions of $\uu$ are to be considered nuisance parameters.  We constrain ourselves here to the case where the dimension $\nx$ of our dataset summary $\xx$ is the same as that of $\uu$; that is, $\nx = \nunknown$.  We hope to explore the extension to $\nx \geq \nunknown$ in future work (see Section \ref{section limitations and future work}).  A pure normalizing flow to describe the distribution of $\xx$, given parameter $\uu$ is simply a neural net mapping $\NN: \RR^{\nx+\nunknown} \mapsto \RR^{\nx}$, with the Jacobian determinant computed only versus the first $\nx$ inputs.  
	
	We propose to decompose this into two separate neural nets, a pivoting net $\zpivot$ and a nuisance net $\znuis$:
	\begin{equation}
		\begin{split}
			\zpivot(\xx; \interest(\uu)) &: \RR^{\nx+1} \to \RR, \\
			\znuis(\xx; \uu)&: \RR^{\nx + \nunknown} \to \RR^{\nx-1}. 
		\end{split}
		\label{equation decomposing}
	\end{equation}
	
	The key point here is that the first net $\zpivot$ is not aware of all parameters $\uu$, but only of the scalar interest function of them $\interest(\uu)$.  The two outputs are concatenated,
	\begin{align*}
		\zz = \begin{pmatrix}
			\zpivot \\ 
			\znuis
		\end{pmatrix},
	\end{align*}
	so that $\zz \in \RR^{\nx}$.

	From here, the same normalizing flow training outlined in Section \ref{section normalizing flows} is followed, to coerce $\zz$ to follow a standard normal distribution, by maximizing the sum across (\ref{equation loss}).  Under a few assumptions about the nature of $\interest$ and of the broader sampling distribution, $\Zpivot | \uu$ will be, \textit{on average across the training distribution of $\uu$}, as close as possible to standard normal (in terms of KL-divergence, see Section \ref{section proof}), while the mapping $\zpivot$ required to generate it requires knowledge of $\uu$ only through $\interest(\uu)$.
	
	\subsection{One further constraint}
	
	We also add one further constraint,
	\begin{align}
		\label{equation constrain negative zp by psi}
		\frac{\partial \zpivot}{\partial \interest} < 0.
	\end{align}
	This constraint helps to prevent $\zpivot$ from flipping orientation and/or folding with respect to $\interest$ (which would otherwise be technically possible, as long as $\znuis$ compensates to keep the entire mapping a diffeomorphism).  It also helps to encourage a clean alignment of $\zpivot$ with the underlying nuisance-free component of the $\interest$-action on the data (see, for example, Appendix \ref{section folding in m}), thereby improving our argument that our architecture should provide reasonable power.
	
	This constraint is enforced in our architecture by adding to the loss
	\begin{align*}
		\kappa \max(\partial \zpivot / \partial \interest, 0),
	\end{align*}
	with $\kappa=100$ chosen large enough to have a large impact at small positive $\partial \zpivot / \partial \interest$, without causing numerical overflows.
	
	\subsection{Obtaining $p$-values from $\zpivot$}
	\label{section p-values}
	
	We convert $\zpivot$ to a one-tailed $p$-value $v$ by the standard normal CDF, $v = \Phi(\zpivot)$.  In Section \ref{section near-uniform}, we show that (under certain assumptions) $\zpivot$ will be (on average across the training distribution of $\uu$) as close to standard normal as the model permits (minimal in KL-divergence from normal).  Equivalently, the average KL divergence between $v$ and uniform is minimized.  These are then converted to equal-tailed two-tailed $p$-values by $2 \min(v, 1-v)$.
	
	\subsection{An example: the one-sample $t$-test}
	\label{section t test example outline}
	
	As a simple example, we might want to reproduce the one-sample $t$-test.  In this case, our parameters $\uu = (\mu, \sigma)$ represent the population mean and standard deviation, with interest parameter $\interest(\uu) = \mu$.  Our statistics would be the sample mean and standard deviation $\xx = (m, s)$.  We would train simultaneously
	\begin{align*}
		\zpivot(m, s, \mu) &: \RR^3 \mapsto \RR, \\
		\znuis(m, s, \mu, \sigma) &: \RR^4 \mapsto \RR^1,
	\end{align*}
	and in our loss function, $\tfrac{\partial \zz}{\partial \xx} = \tfrac{\partial (\zpivot, \znuis)}{\partial (m, s)}$.
	
	We may then obtain a one-tailed $p$-value $v$ for a given sample $(m, s)$ against null hypothesis $H_0: \mu = \mu_0$, by $v = \Phi(\zpivot(m, s, \mu_0))$ and a two-tailed $p$-value by $2\min(v, 1-v)$.

	\section{Formal properties}
	\label{section proof}
	
	\subsection{Assumptions}
	
	Let $\XX|\uu$ denote the random vector of statistics used by the method, with realized value $\xx \in \RR^{\nx}$. We train our model over a large set of parameter values $\uu\in\outerTheta\subseteq \RR^{\nunknown}$, and make inferential claims on a smaller region $\uu\in\Theta_{\mathrm{inner}}\subseteq\outerTheta$ (see Section \ref{section param sampling}). All assumptions below are assumed to hold across $\outerTheta$.
	
	\subsubsection{Suitable models}
	\label{section suitable models}
	
	We constrain our attention to a certain subset of models, for which we believe this architecture in its current form is most promising.  We first assume that the statistic $\XX$ has the same dimensionality as the parameter $\uu$, that is $\nx = \nunknown$.
	
	We further assume that the interest parameter $\interest(\uu)$ is a clean, unfolded coordinate in $\uu$; concretely that $\uu$ can be smoothly and one-to-one reparameterized as 
	\begin{align*}
		\uu\leftrightarrow(\interest,\llambda),
		\qquad
		\llambda\in\RR^{\nunknown-1},
	\end{align*}
	and in this section we write $\uu=(\interest, \llambda)$, with $\interest = \interest(\uu)$.
	
	In particular, we developed our architecture with models in mind of the form
	\begin{align*}
		\XX=\uu+\mathbf{E}_\uu,
		\qquad
		\uu = (\interest, \llambda) \in \RR \times \RR^{\nunknown - 1},
	\end{align*}
	for error $\mathbf{E}_\uu$ that is relatively concentrated and varies smoothly, and slowly, in $\uu$.  We do not at present restrict ourselves exclusively to these models, but our arguments regarding power in Section \ref{section power} in particular will be more easily understood in this context.
	
	In particular, we will assume not only that $\nx = \nunknown$, but also that the transport of the density $p(\xx|\uu)$, induced by perturbations of $\uu$, is full-rank.  That is to say, at each $\uu$, perturbing each dimension of $\uu$ transports the probability mass in a direction linearly independent of that induced by perturbing any other dimension.  As shorthand, we will refer to such dimension-matched and full-rank-transported models as ``\textbf{full-rank models}''.
	
	\subsubsection{Standard regularity assumptions}
	
	We assume that $\XX|\uu$ has a smooth density $p(\xx|\uu)$ $\uu\in\outerTheta$.  We work on a single regular coordinate patch of the model, excluding regions where the model becomes locally singular, folds back on itself, or wraps around so that the same local model is represented by multiple parameter values.
	
	Denoting the log likelihood by $\ell(\uu;\xx)=\log p(\xx|\uu)$, we assume that the score $s(\uu;\xx)=\nabla_\uu \ell(\uu;\xx)$, exists with finite variance. The Fisher information, $I(\uu)=\operatorname{Var}_\uu{s(\uu;\XX)}$, is assumed to exist and have full rank throughout the region considered.
	
	For the formal arguments below, we will additionally assume that the learned transformation is one-to-one on the relevant support (and defines a valid change of variables).  Our present implementation encourages local invertibility through the Jacobian penalty, but does not guarantee global invertibility, as discussed in Section \ref{section invertibility and tractability}.
	
	\subsection{Near-uniformity (by KL) of $p$-values (type I error)}
	\label{section near-uniform}
	
	In the limit of infinite samples, maximizing the likelihood of our constrained normalizing flow has the effect of minimizing the Kullback-Leibler (KL) divergence between our $\zz$ and the normal distribution $\Normal$, averaged across the distribution used to sample $\ttheta$ during training.
	
	Let $\XX$ be the random variable that instantiates into individual samples $\xx$ and
	\begin{align*}
		\Zpivot &= \zpivot(\XX), \\
		\Znuis &= \znuis(\XX) 
	\end{align*}  
	and let $\Normal_d \sim \Normal(\mathbf 0, \mathbf I_d)$ represent a $d$-dimensional standard normal variable. 
	
	By the Kullback-Leibler chain rule, we have:
	\begin{align*}
		\E_\uu \KL((\Zpivot, \Znuis)\|\Normal_{\nx})
		= \E_{\uu} \KL(\Zpivot\|\Normal_1) + \E_{\uu, \Zpivot} \KL(\Znuis|\Zpivot\|\Normal_{\nx - 1}).
	\end{align*}

	Assuming that $\znuis$ is sufficiently flexible to Gaussianize the conditional distribution $\Znuis | \Zpivot$, the final term here can be made (approximately) zero.  In this case, fitting our model, finds the model that minimizes the first term: $\E_\uu \KL(\Zpivot\|\Normal_1)$.  (Of course, our neural network is probably not sufficiently flexible to make the final term exactly zero, but we assume it to be sufficiently flexible to make it negligible.)
	
	Furthermore, KL divergence is invariant under a bijective change of variables, such as conversion of our standard normal $\Zpivot$ into a one-tailed $p$-value, $V = \Phi(\Zpivot)$.  This means that our $p$-values are optimized to have a minimum average KL divergence, 
	\begin{align}
		\label{p values KL}
		\E_\uu \KL(V\|U),
	\end{align} 
	versus a uniform distribution, $U$.  (Note that we are using the random variable $V$ here to refer to a $p$-value to avoid confusion with our densities $p$ and $q$.)

	\subsubsection{Suitability of KL optimization}
	
	\textit{If} there truly exists a perfect pivot, we expect our KL divergence to be minimized at the point where $\zpivot$ is indeed pivotal.  However, in the case where one does not exist, the model must make trade-offs.  In this case KL is a somewhat unnatural loss function.  Minimizing KL divergence does not focus on the sort of tail-specific calibration, or worst-case guarantees that we would ideally want for $p$-values.    
	
	Nevertheless, it has some nice properties in that direction.  With a uniform target $U$, we have $\KL(V\|U) = \int_0^1p(v) \log p(v) \dd v$; for $V$ close to uniform, we can expand $p \log p$ around $p=1$,
	\begin{align*}
		p\log p = (p - 1) + \frac{(p-1)^2}{2} - \frac{(p-1)^3}{6} + \cdots,
	\end{align*}
	and given that $\int_0^1 (p(v) - 1) \dd v = 0$ when $V \in [0, 1]$, we have
	\begin{align*}
		\KL(V\|U)  
		\approx \frac{1}{2} \int_0^1 \left(p(v) - 1\right)^2 \dd v
		= \tfrac{1}{2}\|p(v) - 1\|_2^2.
	\end{align*}
	That is, when $V$ is close to uniform, $\KL$ will trade improvements at one part of the distribution for deviations at another according to the $L_2$ distance between the true \textbf{density} and uniform.
	
	When generating $p$-values, we are most interested in how the \textit{cumulative} distributions compare; unfortunately, relatively small deviations in \textit{density} can accumulate across $v$ into relatively large discrepancies in cumulative distribution.  On the other hand, our interest is particularly in comparing these cumulative distributions \textit{in the tails}.  In the tails, there is much less space for small discrepancies in density to accumulate; the only way to drastically inflate the error rate at $\alpha = 0.05$ is to squash a considerably increased density of $p$-values into the small space $V < 0.05$, which \textit{would} be detected by the Kullback-Leibler divergence.
	
	So, while an ideal method might specifically target CDF uniformity in the tails, optimizing the Kullback-Leibler divergence should nonetheless give a result whose tail CDF is nudged into the right ballpark.  Nonetheless, this is still \textit{on-average} near-pivotal.  It must be stressed that this leaves open a gap for severely non-pivotal behaviour at some parameter values to be traded off against consequently improved behaviour at others.  This is particularly possible at low-probability parameter values, where very poor calibration may nonetheless have a very small impact on the loss.  On the other hand, \textit{if} a true pivot exists, we have good reason to expect that our model will find it.
	
	\subsection{Power (type II error)}
	\label{section power}
	
	Identifying a pivot does not in itself guarantee that that pivot contains any information whatsoever about the interest parameter $\interest$.  Unlike the GAN approach from \cite{louppe2017learning}, our net makes no explicit attempt to retain $\interest$ information in our pivot.  Particularly in cases where $\nx > \nunknown$ (but not limited to these), there may be many ancillary statistics that contain information neither about $\llambda$ nor $\interest$.  However, in the full-rank models (defined in Section \ref{section suitable models}) that we are focussed on, we believe that in general, identifying a pivot \textit{does} recover a useful test statistic.  We make this argument by analogy to the profile likelihood ratio (LR) and its gradient, the efficient score.  
	
	The profile LR achieves excellent power on nuisance parameter problems by starting with the most powerful simple-vs-simple test (the plain LR) and removing $\llambda$ by choosing it at its optimal value.  This has the effect of narrowing the LR for any parts of the data that both $\interest$ and $\llambda$ could equally well explain.  The result is that any contribution of $\interest$ that equally could have come from $\llambda$ is absorbed away in the optimisation, leaving only the pure $\interest$-information behind.  In our model, $\znuis$ has a similar role: it absorbs the nuisance directions, so that $\zpivot$ contains only pure $\interest$-information that cannot be mimicked by changing $\llambda$.  In a ``full-rank'' model (Section \ref{section suitable models}), there are no redundant dimensions available for $\zpivot$ to inhabit other than this one.
	
	We make this more concrete below by reference first to transformation models (in which this effect is quite clean) in Section \ref{section transformation models main text}, before attempting to extend this logic  to a broad subset of non-transformation models, first somewhat heuristically in Section \ref{section near transformation models} and then more concretely based on asymptotics in Section \ref{section lan}.
	
	\subsubsection{Transformation models}
	\label{section transformation models main text}
	
	We can make this link between $\zpivot$ and the profile LR quite tangible in transformation models, where the parameters $\uu$ and corresponding data $\XX$ can be indexed over by the action of a group $G$ with full rank ($\dim G = \nunknown$).  This case covers a huge range of classical examples, such as inference on means and standard deviations of normal models, differences of means and ratios of variances in normal models, linear regression, the ratio of two exponential rates, etc.
	
	In Appendix \ref{section transformation models}, we show how the most natural $\zpivot$ would also be (close to) profile LR sufficient; that is, the profile LR would depend on $\XX$ (almost) only via $\zpivot(\XX; \interest)$.  We argue this first by summarizing previous work showing (Appendix \ref{section maximal invariant}) that in such models, the profile LR must be a function of the maximal invariant of $(\xx, \interest)$ under the action of $g \in G$.  We further argue (Appendix \ref{section zp to maxinv}) that the most natural destination for $\zpivot$ is the same maximal invariant, with the only loss-minimizing alternative being highly globally organized deviations that 
	\begin{enumerate}
		\item are neither encouraged nor discouraged by our loss function,
		\item tend to average out to be very small across a random initialisation, and 
		\item occupy such a small volume in the space of possibilities, that random steps along these directions under stochastic gradient descent are expected to be negligible.
	\end{enumerate}
	In Section \ref{section results toy problem}, we present results of an experiment to demonstrate this effect.  We also show in Appendix \ref{section maxinv power}, that small deviations of this kind would not degrade power significantly.
	
	\subsubsection{Near-transformation models}
	\label{section near transformation models}
	
	In general, non-transformation models have no maximal invariant to hang these arguments on.  However, the core intuition behind the expected power performance in transformation models is: if $\znuis$ absorbs the nuisance directions, then what remains is the pure information about $\interest$, unpolluted by $\llambda$.  We will argue here that the same intuition may be applied to a broad class of non-transformation models also.
	
	At the infinitesimal level, the analogy to transformation models is strong: assuming (as we did with our transformation models) that each parameter transports the probability mass in a linearly independent direction, there is a quotient direction that contains the ``pure $\interest$ information'', after removal of any information which \textit{could} have been accounted for by $\llambda$.  And, since movement induced by $\llambda$ produces, to first order, no displacement along this local quotient direction, it is \textit{locally} pivotal with respect to $\llambda$ and thereby, \textit{locally} appealing to $\zpivot$.
	
	However, \textit{globally}, $\zpivot$ must stitch together these local quotient directions.  Transformation models are the special case where these directions are mutually consistent; each is a local view on the same global coordinate, the maximal invariant.  In general models, the local quotient directions may be inconsistent with each other and the learned $\zpivot$ is there best viewed as a smooth global compromise, with its alignment to the local quotient directions depending on the regions of parameter space emphasized by the training distribution and likelihood.  Formulating a condition under which these local invariants are ``consistent enough'' is a work in progress; such models we refer to here informally as ``near-transformation models''.
	
	As with type I error, this does again leave open the possibility for the architecture to trade poor power in one region of parameter space (particularly one with lower weighting) for improved power at another.  There is no specific component to the loss that rewards an informative pivot and, as such, we will again rely on simulation of the resulting architecture, to understand its properties in this regard.
	
	This argument can be made much more concrete as sample size increases:
	
	\subsubsection{The local and asymptotic case}
	\label{section lan}
	
	Treating the $\nunknown$-dimensional $\xx$ supplied to the architecture as the observed data, suppose that its distribution is locally asymptotically normal (LAN).  This occurs, for example, when $\xx$ is an efficient estimate of $\uu$ based on a growing underlying dataset.  However the class of asymptotically normal estimators for which this holds is far broader.
	
	As the underlying sample size increases, the local region of $\uu$ that is relevant for a given sample $\xx$ shrinks, and on that shrinking local scale, the likelihood loses its higher-order shape, converging effectively to a Gaussian shift experiment.  We explore such an experiment in Appendix \ref{section gaussian location}, and summarize the conclusions here, adapted as they relate to this local asymptotic normal (LAN) case:
	
	\begin{enumerate}
		\item Local to parameter $\uu_0$, the likelihood function $\ell(\uu)$ is approximately quadratic,
		\begin{align*}
			-2(\ell(\uu) - \ellhat) \approx (\uu - \uuhat)^\top \iinfo (\uu - \uuhat), \qquad \uuhat = \uu_0 + \iinfo^{-1} \sscore,
		\end{align*}
		where $\sscore = \sscore(\uu_0;\xx)$ and $\iinfo = \iinfo(\uu_0)$ are the score and Fisher information (respectively), evaluated at fixed $\uu_0$ close to $\uu$, and $\ellhat = \ell(\uuhat)$ is the constant likelihood at the peak of the quadratic.
		
		\item Here $\uuhat = \uu_0 + \iinfo(\uu_0)^{-1}\sscore(\uu_0; \xx)$ is the random variable in our local asymptotic Gaussian-shift experiment, $\uuhat \dot\sim \Normal(\uu, \iinfo^{-1})$.  But it also plays the role of a simple conversion of $\xx$ into a more intuitive form, which locates the $\uu$ at the turning point of the quadratic, so that it can be expressed as a perfect square.  Computing $\uuhat$ this way is also the only adaptation necessary to convert the Gaussian location experiment in Appendix \ref{section gaussian location} into its LAN equivalent here.
		
		\item Profiling this quadratic absorbs any information that can be explained by $\llambda$, to leave a one-dimensional quadratic profile likelihood $\ell_p$ that has the same form as the marginal likelihood for $\interest$: it is centred at the same $\interesthat$ as the $\interest$- dimension of the full likelihood $\ell$, but with potentially reduced curvature, capturing information lost to $\llambda$.  Subtracting the maximum, $\ell_p(\interesthat)$ (a constant in $\interest$), gives the log profile LR $\Lambda$,
		\begin{align*}
			&-2\log \Lambda(\interest) = \\
			&-2(\ell_p(\interest) - \ellhat_p) \approx (\interest - \interesthat)^2 \effinfo,
		\end{align*}
		where $\interesthat$ is the $\interest$- (first-) element of $\uuhat$,
		\begin{align*}
			\interesthat = \uuhat_\interest = \interest_0 + \effscore / \effinfo,
		\end{align*}
		$\effscore$ is the efficient score, $\effinfo$ is the efficient information
		\begin{align*}
			\effinfo^{-1} = (\iinfo^{-1})_{\interest\interest},
		\end{align*}
		and $\ellhat_p = \ell_p(\interesthat)$ is the constant profile likelihood value at the peak of the quadratic.
		
		\item The maximal invariant of this local Gaussian location experiment is
		\begin{align*}
			\pivotlocal = \interesthat - \interest.
		\end{align*}
	\end{enumerate}
	
	We can see that the data enters the maximal invariant (the natural target for $\zpivot$, see Appendix \ref{section zp to maxinv}) only through the efficient score: an object known to be highly informative about $\interest$.  Furthermore (and as expected), the powerful profile LR is a function of this maximal invariant.
	
	What is more, after a simple local rescale, this maximal invariant is known to converge in distribution, $\sqrt{\effinfo} \pivotlocal \xrightarrow{d} \Normal(0, 1)$, making it asymptotically pivotal and so an ideal target for $\zpivot$.  (The $\chi^2$ limiting distribution of $-2\log \Lambda$ results directly from this.)  This convergence \textit{in part} reflects the shrinking local region over which the nuisance effects must be unified.  The same benefit can be expected for $\zpivot$: consistency between the local maximal invariants is required over a progressively smaller neighbourhood, making the job of combining them into a global $\zpivot$ progressively easier.
	
	The efficient score, profile LR and local maximal invariants all then embody precisely the intuition we are aiming at: that removal of $\llambda$-information distils the pure $\interest$-information.  In the LAN limit, all three draw from the same local information about $\interest$, and we might expect $\zpivot$ to converge to their performance in terms of power.

	\section{Further considerations}
	\label{section further considerations}
	
	In Section \ref{section invariances} we discuss how we incorporate invariances into our model and in Section \ref{section known params} we further add extra ``known values'' (like sample size) to allow our model to work over a wider range of inputs.  In Section \ref{section param sampling}, we discuss how to sample parameters so that we have confidence in our $p$-values.  Finally in Section \ref{section neural network design}, we discuss small tweaks to the neural network design that we have found anecdotally to be helpful.

	\subsection{Incorporating invariances}
	\label{section invariances}
	
	In many cases, there are invariances in the structure of our model that we can exploit.  In the one-sample $t$-test example (Section \ref{section t test example outline}), we expect the same behaviour irrespective of location and scale.  If we can incorporate these invariances into our neural network mappings, we can reduce their complexity (reducing their dimensionality), while simultaneously expanding the domain over which they can produce valid $p$-values (across all locations on a given orbit; in many cases this makes the domain infinite in one or more dimensions).
			
	Suppose that the model we wish to fit is invariant (up to the corresponding Jacobian factor $|J_g|$) under the action of a transformation group $G$; specifically $\forall g \in G$,
	\begin{align*}
		p(\xx|\uu) &= p(g\cdot \xx|g\cdot \uu) |J_g|, \\
		\interest(g\cdot\uu) &= g \cdot \interest(\uu).
	\end{align*}
	where $|J_g|$ represents the absolute Jacobian determinant of the mapping $g$, and $G$ has $\nG$ dimensions.
	
	We introduce a canonicalization rule $\sx \coloneqq \s(\xx): \xset \mapsto G$, which chooses a mapping, based on the statistics $\xx \in \xset$, onto a single canonical representative from the orbit on which $\xx$ lies.  In doing so, it also reduces the dimensionality of $\sx \cdot \xx$ to $\nx - \nG$.  Since the canonicalization rule is chosen based on $\xx$, no dimensions are lost from $\uu \mapsto \s \cdot \uu$ and its dimensionality remains $\nunknown$.
	
	Then we replace each of our normalizing flows by the canonicalization transform followed by a network with smaller input dimension:
		\begin{equation}
		\begin{split}
			\zpivot(\sx \cdot \xx; \interest(\sx \cdot \uu)) &: \RR^{(\nx-\nG)+1} \to \RR, \\
			\znuis(\sx \cdot \xx; \sx \cdot \uu)&: \RR^{(\nx - \nG) + \nunknown} \to \RR^{\nx-1}. 
		\end{split}
		\label{equation decomposing invariant}
	\end{equation}
	Critically, when the Jacobian is computed for the loss in (\ref{equation loss}), it is still computed with respect to the full (pre-canonicalization) $\xx$, by chain rule through $\sx$,
	\begin{align*}
		\frac{\partial \zz}{\partial \xx} = \frac{\partial \zz}{(\sx \cdot \xx, \sx \cdot \uu)} \frac{\partial (\sx \cdot \xx, \sx \cdot \uu)}{\partial \xx}.
	\end{align*}
	
	One might wonder whether our Jacobian $\tfrac{\partial \zz}{\partial \xx}$ can still be nonsingular, given that we first transform $\xx \in \RR^{\nx}$ to a lower dimensional $\sx \cdot \xx \in \RR^{\nx - \nG}$.  The key intuition is that some of the gradient with respect to $\xx$ is also transferred via $\sx \cdot \uu$, which now also depends on $\xx$.  We must have $\nG \leq \nx$ by definition and we have $\nx = \nunknown$ by assumption (Section \ref{section decomposing}), so there will always be at least enough dimensions in our neural network for a full-rank Jacobian to be possible.
	
	Again the $t$-test represents a simple example: we can capture scale invariance by $\sx \cdot \xx = (m/s)$ and $\sx \cdot \uu = (\tfrac{\mu}{s}, \tfrac{\sigma}{s})$.  We can even reduce $\xx$ to zero dimensions by further incorporating location invariance: $\sx \cdot \xx = ()$ and $\sx \cdot \uu = (\tfrac{\mu - m}{s}, \tfrac{\sigma}{s})$.  Even though there are no inputs to our net that directly represent $\xx$, the Jacobian with respect to $(m, s)$ is still transferred via $\sx \cdot \uu$ by 
	\begin{align*}
		\frac{\partial \zz}{\partial \xx} = 
		\frac{\partial \zz}{\partial (\tfrac{\mu - m}{s}, \tfrac{\sigma}{s})} \frac{\partial (\tfrac{\mu - m}{s}, \tfrac{\sigma}{s})}{\partial (m, s)}.
	\end{align*}  	
	
	We reproduce exactly this example in practice in Section \ref{section t-test methodology} and present in Section \ref{section behrens fisher} promising numerical results from an example where $\nx = \nunknown = 3$ and $\nG = 2$, so that here also the Jacobian \textit{must} flow via the parameters.

	\subsection{Conditioning on further known values}
	\label{section known params}
	
	As described in Section \ref{conditional flows subsection}, adding extra conditioning variables to our normalizing flows can be achieved simply by adding those conditioning variables as further inputs to the network.  We use this to allow further ``known values'' $\kk$ (of dimension $\nknown$) to be added to the model.  Again referring to the $t$-test example, we might take the sample size $\nu = n$ as a known value, so that our net can model the $t$-test problem over a range of sample sizes.  
	
	Then our full normalizing flow decomposition (\ref{equation decomposing}) is expanded to
		\begin{equation}
		\begin{split}
			\zpivot(\xx; \interest(\uu), \kk) &: \RR^{\nx+1+\nknown} \to \RR, \\
			\znuis(\xx; \uu, \kk)&: \RR^{\nx + \nunknown + \nknown} \to \RR^{\nx-1} 
		\end{split}
		\label{equation decomposing known}
	\end{equation}
	and likewise for the $G$-invariant version in (\ref{equation decomposing invariant}).

	\subsection{Sampling an appropriate range of parameters for training}
	\label{section param sampling}
	
	If $\zpivot$ is to learn how to pivot, then every $\xx$ that is likely to arise from a given set of parameters $\ttheta_1$ must compete with the \textit{same} $\xx$ value arising from every other set of parameters $\ttheta_2$ with the same $\interest(\ttheta_1) = \interest(\ttheta_2)$ that are ``likely'' to give rise to it.  Therefore, the choice of parameter samples $\paramset$ to be included in training the net is rather important.
	
	Define $\xtarget \subseteq \xset$ to be the set of all $\xx \in \xset$ for which we want to be able to generate valid $p$-values.  Further, for any $\paramset_a \subseteq \paramset$, define $\xset_{\paramset_a} \subseteq \xset$ to be the set of all $\xx$ that have non-negligible probability of arising from one or more $\uu \in \paramset_a$.  Parameter samples $\uu$ for testing and training our net are then drawn respectively from:
	\begin{enumerate}
		\item $\innerTheta = \{\uu : \xset_{\{\uu\}} \cap \xtarget \neq \varnothing\}$, an ``inner'' set of parameters, which represent all $\uu$ that might generate values in $\xtarget$ --- we want our model to generate valid $p$-values for all $\uu \in \innerTheta$ and use this for testing;
		\item $\outerTheta = \{\uu : \xset_{\{\uu\}} \cap \xset_{\innerTheta} \neq \varnothing\}$, an ``outer'' set of parameters, which represent all $\uu$ that might generate $\xx$ that coincide with any of those generated by $\innerTheta$ --- we train our model on $\uu \in \outerTheta$.
	\end{enumerate}
	
	Within the respective sets $\innerTheta$ and $\outerTheta$, we would like to sample our parameters according to the Jeffreys prior $p(\uu) \propto \sqrt{\det(I(\uu))}$.  Since the Fisher information $I$ is somewhat expensive to estimate in our setup, we instead note that for maximum likelihood estimators $\uuhat$ of $\uu$, $\Cov(\uuhat|\uu) \approx I(\uu)^{-1}$ and approximate $I$ from the covariance matrix of a series of estimates computed from samples drawn from $\uu$.  We then use the square root determinant of these approximate $I$ to train one \textit{denormalizing} flow (Section \ref{section importance sampling flows}) each to model $\innerTheta$ and $\outerTheta$ and allow rapid sampling from them.  
	
	We further use the same mapping from data to parameter estimates, $\xx \mapsto \uuhat$, to define $\xtarget$ as a bounding box $\boundingbox$ around these estimates.  We refer to this bounding box as the ``estimates box''.  The procedure for achieving all of the above is currently a little \textit{ad hoc}, and we hope to refine it in future work.  It is described in Appendix \ref{parameter sampling appendix}.
	
	\subsection{Neural Network Design}
	\label{section neural network design}
	
	A neural network typically consists of simple differentiable functions ``stacked on top of each other''.  The most common format for each ``layer'' $\NNi{i}$ is the fully-connected layer, which looks rather like a series of parallel generalized linear models:
	\begin{align}
		\label{equation fully connected}
		\NNi{i} = \sigma(\mathbf{W}^{(i)} \xx + \mathbf{b}^{(i)}),
	\end{align}
	for some non-linearity $\sigma$ and with the matrix $\mathbf{W}$ of weights and vector $\mathbf{b}$ of biases representing the learnable parameters $\phi^{(i)}$ for that layer.  Then the entire neural network is made of the composition of a series of these layers:
	\begin{align*}
		\NN = \NNi{n} \circ \NNi{n-1} \circ \cdots \circ \NNi{1}.
	\end{align*}
	In NeuralCIs, our input and output layers are based on the classic fully connected layer, but with all intermediate (``hidden'') layers incorporating an extra multiplication and addition with a standardisation in-between:
	\begin{align}
		\NNi{i} = \sigma\left(\xx \odot (\mathbf{W}^{(i)} \xx + \mathbf{b}^{(i)})\right) + \xx,  \label{equation multiplyer layer}
	\end{align}
	where $\odot$ represents the Hadamard product.  Our $\sigma$ now represents a ``layer-norm'', which standardizes the elements of its input vector by their combined mean and standard deviation.
	
	The key intuition behind the incorporation of the extra Hadamard product is that it should allow each input more easily to moderate (``interact with'') the effect of another.  In a classic fully-connected layer (\ref{equation fully connected}), a variable can moderate the effect of another only by means of cancellations of effects across different neurons.  In our design (\ref{equation multiplyer layer}), a given input can directly moderate the effect of any other input that is directed into its output path by $\mathbf W^{(i)}$.
	
	Furthermore, the extra multiplication and layer-norm provide a clean, simple non-linearity, thereby removing the need for an artificially chosen $\sigma$.  The standardization helps to keep values from exploding or vanishing as they are propagated, as well as gradients as they are propagated backwards by chain rule.  The final addition similarly helps to stop the gradients from dying out as they are propagated backwards, by providing a form of ``skip connection''.
		
	Anecdotally, this has performed somewhat better for our models.  Removing the multiplication and instead adding an ELU activation in our tests in Section \ref{section experiments} provided slightly less accurate test sizes across the board (re-checked at various stages in the project).  For Behrens-Fisher, standard deviation of test sizes at $\alpha=0.05$ in our latest simulation (across 10000 parameter samples, each with one million $p$-values) was 0.0032 with multiplication vs 0.0035 with the ELU.  For partial biserial correlation (the no-comparison run in Section \ref{section methodology neuralcis alone}, again 10000 samples each of one million $p$-values), these were 0.0033 (multiplication) vs 0.0038 (ELU).  We might expect the effect to be more pronounced the larger the input dimension and the more complex the interaction structure between those inputs.  But we leave a full analysis of this for a later paper.
	
	Inputs to the nets are transformed as appropriate for the type of variable (log for SDs, atanh for correlations) and rescaled to focus their distribution very roughly onto $[-1, 1]$.  Our model uses classic fully-connected layers at input and output to expand/shrink the dimension to/from a hidden dimension of 50.  These layers have no non-linearity ($\sigma(x) = x$).  Between these two layers are sandwiched 7 hidden layers following (\ref{equation multiplyer layer}) with dimension 50.  The full model design has not yet been fully optimized; we leave that also to future work.
	
	Each net was optimized using Keras' Nadam\footnote{Adam with Nesterov gradients.} optimizer, with an initial learning rate of 0.0025, decayed by a ratio of 0.9 after 8 epochs with no improvement in training loss, to a minimum learning rate of $10^{-6}$.  The batch size was 1024; the $(\zpivot, \znuis)$ nets were (except the toy problem in Section \ref{section toy problem}) trained with 1000 epochs of 200 steps, and all other nets with 500 epochs of 100 steps.
			
	\section{Experimental Methodology}
	\label{section experiments}
	
	To evaluate the performance of NeuralCIs, we ran Monte Carlo simulations to compare to standard approaches on three classic statistics problems with nuisance parameters: normal mean with unknown variance, the Behrens-Fisher problem and partial biserial correlations.  We further ran a toy simulation to illustrate the convergence of $\zpivot$ to the maximal invariant.
	
	In each simulation, we ensure that the same random samples are used for each method (for example by using stateless random number generators) so that the data is truly paired.  Power values are size-adjusted: that is, they are computed based on thresholding each $p$-value at an ``idealized'' value that corrects for the test being conservative or liberal.  Without such a correction, a very liberal test can easily achieve higher power, without having detected anything more of substance.  
	
	This size adjustment is achieved by adding a further simulation that (i) draws samples from the alternative hypothesis instead of the null, (ii) computes a $p$-value for each such sample against a null hypothesis equal to this alternative, and (iii) finds the $\alpha$ quantile of these $p$-values and uses this as the threshold, rather than $\alpha$.  If the size of the particular method is correct, this should simply yield $\alpha$ as the threshold, but if the test is liberal, it will result in a lower threshold and correspondingly lower power, which represents the power the test \textit{would} achieve, \textit{if} the test were correctly calibrated and therefore right-sized.
	
	\subsection{$t$-Test}
	\label{section t-test methodology}
	
	According to Section \ref{section transformation models main text}, we expect $\zpivot$ to align with the maximal invariant $\pivot$ of $(\xx, \interest)$ in a transformation model.  As an example, we fit the classical one-sample $t$-test, described in Section \ref{section t test example outline}.  To test this, we compare one-tailed $p$-values for a range of samples via the classical $t$-test with those generated via NeuralCIs.  We fit two different NeuralCIs models for this.  The first does not account for location and scale invariance (we will refer to this as the ``non-canonicalized'' model); the second (the ``canonicalized'' model) does, as described in Section \ref{section invariances}.  
	
	In the non-canonicalized model, we draw sample sizes $n \in [3, 100]$ and aim to be able to correctly generate $p$-values for parameter estimates $\hat \mu = m \in [-3, 3]$ and $\hat \sigma = s \in [0.333, 3]$.  To be able to generate well-calibrated $p$-values for these samples, we might hope that any $(\mu, \sigma)$ that \textit{could} generate  such samples would generate uniform $p$-values across all samples, even those outside of the estimates box.  This is the basis of our definition of $\innerTheta$ in Section \ref{section param sampling}.  Our sampling scheme at present does not successfully reach to the very edge of $\innerTheta$, see Appendix \ref{section ttest inner theta appendix}, so in the main text of this first version of the paper, we present results for the reduced problem, where $\xx$ are drawn from \textit{parameters} drawn from the estimates box, $\xx \sim p_{\xx}(\xx|\uu)$ with $\uu \in \boundingbox$, rather than the full problem $\uu \in \innerTheta$.  
	
	Specifically, we draw 50000 $\mu \in [-3, 3]$ samples uniformly and 50000 $\sigma \in [0.333, 3]$ and $n \in [3, 100]$ samples log-uniformly.  We then draw one $(m, s)$ pair randomly at each parameter; that is, a total of 50000 samples.  We hope to be able to expand this to the full $(\mu, \sigma) \in \innerTheta$ in a subsequent version of the paper soon.
		
	In the canonicalized model, we hope to obtain valid $p$-values with any $(m, s)$ drawn from any $(\mu, \sigma)$ and with $n \in [3, 100]$.  We test this by drawing 50000 $\mu \in [-100, 100]$ and $\sigma \in [0.01, 100]$ samples; we then draw a sample each from the sampling distributions at these parameters.
		
	\subsection{Behrens-Fisher}
	\label{section behrens fisher}
	
	The Behrens-Fisher problem is a perfect example to test this on.  Data for each of two groups are assumed to have been drawn from one normal distribution per group, each having its own mean ($\mu_1, \mu_2$) and standard deviation ($\sigma_1, \sigma_2$).  The goal is to infer the difference $\Delta = \mu_2 - \mu_1$ between the means of the two normal distributions.  Finding a test statistic for the Behrens-Fisher problem that is near-pivotal with a known reference distribution, while retaining good power, has long been an active area of research: \citet{paul2019review} review and simulate a number of alternatives and ultimately recommend the very widely used approximation developed by \citet{welch1947generalization} in most small-to-moderate sample-size configurations.
	
	We train our model on the parameters $\Delta$, $\sigma_1$ and $\sigma_2$ with statistics $\hat \Delta$, $\hat \sigma_1$ and $\hat \sigma_2$ based on their unbiased and sufficient statistics\footnote{For standard deviations, we use the square root of the unbiased estimator of the variance, and $\hat \Delta = \hat \mu_2 - \hat \mu_1$ is the difference of the two further sufficient statistics: the sample means.}.  We further add the group sample sizes $n_1, n_2 \in [3, 100]$ as ``known values'' as per Section \ref{section known params}.  We model location and scale invariance by subtracting $\hat \Delta$ from both $\Delta$ and $\hat \Delta$ and dividing all parameters and statistics by $\hat \sigma_1$.  This reduces the set of statistics that will enter our net to just one: $\tfrac{\hat \sigma_2}{\hat \sigma_1}$.  We train our network to target valid $p$-values for any $\tfrac{\hat \sigma_2}{\hat \sigma_1} \in [0.333, 3]$.
		
	We aim to select a wide and representative sample of parameters to test the model.  We do not use either of the denormalizing flows that are trained according to Section \ref{section param sampling} and used in training the net, as these might conceal holes in the training data.  Instead we sample a fresh set of 10,000 parameter combinations as described in Table \ref{test params table}, with which we would like to stress-test the invariance logic, by being far outside of the range of parameters encountered within training.
	
	However, as with the $t$-test example in Section \ref{section t-test methodology}, we currently draw samples from \textit{parameters} that have $\sigma_2 / \sigma_1 \in [0.333, 3]$ rather than any parameter that can produce statistics in that range.  Again, this relates to a slight weakness at the very edges in the current parameter sampling scheme, see Appendix \ref{section ttest inner theta appendix}, which we are working to fix.  We hope to expand this to the full $\innerTheta$ in a subsequent revision soon.
	
	\begin{table}[ht]
		\centering
		\begin{tabular}{lll}
			\hline
			Parameter & Distribution & Comment \\
			\hline
			Sample size group 1 & $n_1 \sim \mathrm{LogUnif}(3, 101)$ & Floored (integer 3-100)\\
			Sample size group 2 & $n_2 \sim \mathrm{LogUnif}(3, 101)$ & Floored (integer 3-100) \\
			SD group 1 & $\sigma_1 \sim \mathrm{LogUnif}(0.01, 100)$ &  \\
			SD group 2 & $\sigma_2 \sim \mathrm{LogUnif}(0.333\sigma_1, 3\sigma_1)$ &  \\
			Mean difference & $\mu_2 - \mu_1 \sim \mathrm{Unif}(-100\sigma_1, 100\sigma_1)$ & \\
			\hline
		\end{tabular}
		\caption{Sampling distributions of Behrens-Fisher parameters in our test.}
		\label{test params table}
	\end{table}

	For each parameter sample $\uu = (\Delta, \sigma_1, \sigma_2)$, we further compute a corresponding $\Delta^{\mathrm{power}}$ value that aims \textit{crudely} to target a power drawn uniformly in the range $1 - \beta \in [0.05, 0.90]$, when applied to statistics sampled from the corresponding null parameters $\uu$ at a $p$-value threshold of $\alpha = 0.05$:	
	\begin{align*}
		\Delta^\mathrm{power} &= \Delta \pm \sigma\left(\Phi^{-1}(1 - \tfrac{\alpha}{2}) + \Phi^{-1}(1 - \beta)\right), \\
		\sigma &= \sqrt{\frac{\sigma_1^2}{n_1} + \frac{\sigma_2^2}{n_2}},
	\end{align*}
	where $\Phi^{-1}$ is the standard normal quantile function; where $\pm$ is randomly assigned to addition or subtraction; and where $1 - \beta$ is a target power value uniformly sampled from $[0.05, 0.90]$.
	
	At each sampled $(\Delta, \Delta^{\mathrm{power}}, \sigma_1, \sigma_2, n_1, n_2)$, we compute the following three sets of $p$-values:
	\begin{enumerate}
		\item \label{null p-value sim} At $(\Delta, \sigma_1, \sigma_2, n_1, n_2)$, we draw $10^6$ samples of statistics $(\hat \Delta, \hat \sigma_1, \hat \sigma_2)$, and pass $(\hat \Delta, \hat \sigma_1, \hat \sigma_2, \Delta, n_1, n_2)$ into our model to generate $10^6$ null-hypothesis $p$-values.
		\item To assess power, we also pass the \textit{same} $10^6$ samples into our model as $(\hat \Delta, \hat \sigma_1, \hat \sigma_2, \Delta^{\mathrm{power}}, n_1, n_2)$.
		\item To further size-adjust this power, we repeat step \ref{null p-value sim} with $\Delta^{\mathrm{power}}$ in place of $\Delta$, for a smaller sample of $10^4$ size-adjust $p$-values.
	\end{enumerate}
	We pass the same values across the three steps into the Welch test.  
	
	\subsection{Partial biserial correlation}
	\label{section biparcorr}
	
	As a stress test for the architecture, we also chose to model the partial biserial correlation, where a latent partial correlation between two normal random variables $A \sim \Normal$ and $B \sim \Normal$, conditioned on a third $C \sim \Normal$, is masked by one of the former two variables only being observable through its binarization $\Astar$, where $\Astar \coloneqq 1\{A > \tau\}$ for some threshold value $\tau$.
	
	We model the simplified case, where all normal variables are known to have zero mean and unit standard deviation.  In that case, the parameters to the model are the correlations between the three normal variables $\rho_{AB}, \rho_{AC}, \rho_{BC}$ as well as the binarization threshold $\tau$.  We find it more intuitive to model the threshold as a binarization probability $\pi_{\Astar} = \Phi(-\tau)$.  Importantly, it is also cleaner to model the correlation $\rho_{AB}$ by its partial correlation $\rho_{AB | C}$, so that each of the three correlation parameters can take any value $\rho \in (-1, +1)$ without losing positive definiteness.  	As with our Behrens-Fisher model, we also add the sample size $n$ as a ``known value''.
	
	We choose a set of cheap statistics: the observed (e.g. \textit{point} biserial) correlations between the three observable variables $\hat \rho_{\Astar B}, \hat \rho_{\Astar C}, \hat \rho_{BC}$ along with the observed prevalence in $\Astar$, $\hat \pi_{\Astar}$.  Note that the statistics here are \textbf{not} direct estimates of the parameters, as the correlations with $\Astar$ are point biserial correlations with binary $\Astar$.  They are also not known to be sufficient statistics.  However, under the latent Gaussian threshold model, the prevalence and point-biserial correlations are in one-to-one correspondence with the threshold and latent biserial correlations; see for example \cite{olsson1982polyserial} for the relationship between biserial and point-biserial correlations.
	
	This represents a stress test for two reasons in particular.  Firstly, our statistics do not form a sufficient statistic for the parameter vector of the model.  Secondly, one of these statistics, $\hat \pi_{\Astar}$ is discrete rather than continuous.  Since the entire theory is based on diffeomorphisms on continuous statistics, it raises the interesting question: can our model approximate in the discrete case?  
	
	We compare our approach with two widely used likelihood-ratio (LR) -based approximations: the $\chi^2$ likelihood-ratio test and the bootstrap likelihood-ratio test:
	
	\subsubsection{Likelihood ratio (LR) tests}
	\label{section lr tests}
	
	A $p$-value simulation involving likelihood ratios must find the maximum likelihood estimate (MLE) of the parameters (once constrained to $\rho_{AB|C} = \rho_{AB|C}^{\mathrm{null}}$, once unconstrained) for every one of hundreds to millions of sampled datasets per parameter set that we wish to simulate from.  Any dataset whose MLE lands at the boundary of our space (i.e. correlations close to $\pm 1$, prevalence close to 0 or 1) can be problematic, both for our optimizer and (particularly for directly-on-the-boundary samples) for our interpretation of the likelihood ratio.  We aim to keep the number of boundary cases per parameter to a rate that is negligible, by selecting parameters $\uu$ to simulate that are sufficiently far themselves from the boundary.  We then track and report non-convergence and boundary case rates, allowing each of these to remain in our $p$-value curve, rather than excluding them.
	
	In Appendix \ref{appendix sampling biparcorr params}, we outline a sampling scheme for $\uu$ that aims to keep boundary correlation cases ($|\hat \rho| > 0.99$) and degenerate prevalence cases ($\hat \pi_{\Astar} \in \{0, 1\}$) rare, to the order of approximately $10^{-4}$ or less.  For low $n$, this tends to mean quite a narrow range for our parameters (for $n=20$, we have $\pi_{\Astar} \in (0.39, 0.61)$, $|\rho_{AB|C}| < 0.07$, $|\rho_{AC}| < 0.11$, $|\rho_{BC}| < 0.93$), and this gradually fans out to a fairly broad range for higher $n$ (for $n=100$, $\pi_{\Astar} \in (0.09, 0.91)$, $|\rho_{AB|C}| < 0.72$, $|\rho_{AC}| < 0.72$, $|\rho_{BC}| < 0.98$).  We further pair each $\interest = \rho_{AB | C}$ parameter sample with a corresponding $\rho_{AB | C}^\mathrm{power}$ value, used to compute power against the same sampled statistics with just one extra likelihood optimization; this is sampled to very roughly target a uniform distribution for power $(1 - \beta) \sim \mathrm{Unif}(0.05, 0.90)$, but constrained to the same range.
	
	Prevalences parameterizing the likelihood surface were mapped from $(0, 1)$ to $(-1, 1)$ and all parameters were then atanh-transformed.  In this transformed space, there is quite a wide spread at the boundaries, leading occasionally to very slow convergence for extreme samples (delays, which are then multiplied across many parallel runs).  To speed up convergence, the likelihood surface corresponding to extreme correlation values, $|\hat \rho| > 0.99$ was (in the optimization phase) swapped for a steep downward slope, so that the optimizer would not wander into this region.  The same was applied to the transformed prevalences, i.e. a ramp was applied at $|2\hat \pi_{\Astar}-1| > 0.99$.  This ramp was removed when computing likelihood ratios, but had the effect of clipping extreme values around $\pm 0.99$.  We explore the rates at which samples ended up caught on this fold (``boundary samples'') or with all-equal $\Astar$ values (``degenerate samples'') in Appendix \ref{section convergence}.
	
	\subsubsection{Likelihood ratio $\chi^2$-test}
	
	In the large sample limit, $-2 \log LR$ is asymptotically $\chi^2_1$ distributed.  This then allows for a relatively cheap test to be performed, in which two MLEs are evaluated and converted into an LR and directly into a $p$-value.
	
	Nonetheless, optimizing to find the MLE twice in each case is far more computationally expensive than NeuralCIs, so we have to constrain ourselves to just 1000 parameter samples (compared with 10000 for Behrens Fisher, Section \ref{section behrens fisher}).  As before, we generate one million sampled datasets at each parameter, to produce stable estimates of error rates and power.  (This large number was achieved by running parallelized optimisations on an RTX 3060 Mobile GPU, using the Tensorflow-Probability BFGS optimizer.)  The size-adjust is again based on a smaller sample of ten thousand.
		
	\subsubsection{Bootstrap likelihood ratio test}
	
	An alternative, and quite general approach to converting the profile LR to a $p$-value is to approximate its null distribution by parametric bootstrap.  Samples are generated from our model, with $\rho_{AB|C} = \rho_0$ and nuisance parameters fixed at their constrained MLEs.  (This is the same constrained MLE described in Section \ref{section lr tests}, optimized for the dataset in question.)  For each sample, the profile LR is computed as described in Section \ref{section lr tests}, and the resulting bootstrap LR statistics are used as an approximation to the true null distribution.  This is only an approximation, because we cannot guarantee (and often do not expect) that the null distribution of the LR will be the same across all relevant values of the nuisance parameters.
	
	This is computationally several orders of magnitude more costly than the $\chi^2$ LR test, because each bootstrap sample costs as much as a single $\chi^2$ LR test.  For this comparison, we ran 500 parameter samples, generating 500 $p$-values per parameter sample, each with 2000 bootstrap samples.  We use the same number of samples and simulations again for the size-adjust.
	
	\subsubsection{Performance of NeuralCIs alone}
	\label{section methodology neuralcis alone}

	Comparing NeuralCIs with bootstrap LR imposes a number of constraints, in particular a constrained parameter range (to avoid boundary statistics) and the small number of simulations that can be run (due to computational cost).  Therefore, we will also present false positive rates of NeuralCIs without comparison to another method, with a much larger sample (as with Behrens-Fisher, one million $p$-values at each of 10,000 parameter samples), drawn much closer to the boundary.  
	
	To avoid overemphasizing the boundaries (and giving negligible probability mass to the important zero-correlation centre), we approximate drawing parameters at uniformly sampled angle and \textit{radius} as follows.  First a point is chosen uniformly from the 4D hypercube $[-1, 1)^4$ and turned to a unit vector by dividing by its magnitude.  This ``angle'' is then multiplied by a uniformly sampled magnitude in $[0, 2)$.  The fourth dimension is remapped $(-1, 1) \mapsto (0, 1)$ and the four dimensions are assigned to $\rho_{AB | C}, \rho_{BC}, \rho_{AC}$ and $\pi_{\Astar}$, respectively.  Finally, any correlation greater in absolute value than 0.9 and any  $\pi_{\Astar}$ not in the range $0.1 \leq \pi_{\Astar} \leq 0.9$ is excluded.
	
	\subsection{Toy problem: measuring deviation from maximal invariant}
	\label{section toy problem}
	
	In Appendix \ref{section u dependence marginalized}, we show that a model that is shift-uniform in the nuisance-orbit coordinate $\uuu$ would allow construction of a pivot that can be contaminated by a periodic dependency on $\uuu$ (that is, depend on $\xx$ other than through the maximal invariant).  We make an implicit-bias argument that such a solution is highly unlikely to be discovered when fitting $\zpivot$.
	
	To test this, we further fit a toy problem that has exactly this shift-uniform structure and monitor how closely $\zpivot$ depends on the maximal invariant $\pivot$ at each training epoch.  We fit our architecture to the following two-dimensional model, with scalar parameters $(\interest, \lambda)$:
	\begin{alignat*}{2}
		X_1 &= \interest + X_2 + E_1, &\qquad
		X_2 &= \lambda + E_2, \\
		E_1 &\sim \Normal(0, 1), 
		&\qquad
		E_2 &\sim \mathrm{Unif}(-0.5, 0.5).
	\end{alignat*}

	In this model, the maximal invariant $M_\interest$ and the nuisance orbit coordinate $U$ are
	\begin{align*}
		M_\interest = X_1 - X_2 - \interest,
		\qquad
		U = X_2.
	\end{align*}
	
	Before training, we construct a set of reference values of $m$, $e_2$, $\interest$ and $\lambda$.  For $m_\interest = e_1$, we use 32 evenly spaced quantiles of the normal distribution, divided by their standard deviation to give them exact standard deviation of 1.  For $e_2$, we use a uniformly distributed random sample of 64 values drawn between -0.5 and 0.5.  For parameters we use 16 linearly spaced values each, for $\interest$ in $[-2.2, 2.2]$ and for $\lambda$ in $[-1, 1]$.  
	
	Let $\E_m, \mathrm{Var}_{\interest, \lambda}$ denote means and variances across the respective fixed grids $m, u, \interest, \lambda$ defined above.  After each epoch, we compute $\zpivot$ and from these the following metrics.
	
	\begin{enumerate}
		\item A measure of the proportion of $\Zpivot$-variance contributed by $U$, estimated by the proportion of $\zpivot$-variance not accounted for by $m$:
		\begin{align*}
			\mathrm{U\ proportion} = \max_{\interest, \lambda} \frac{\E_m [\mathrm{Var} (\zpivot|m)]}{\mathrm{Var}(\zpivot)}
		\end{align*}
		\item A measure of how similar the curve mapping $M$ to $\Zpivot$ is across different $(\interest, \lambda)$:
		\begin{align*}
			\mathrm{Mapping\ stability} = \max_{m} \mathrm{Var}_{\interest, \lambda}(\E[\zpivot|m]).
		\end{align*}
		\item A measure of the variance of $\Zpivot$:
		\begin{align*}
			\mathrm{Output\ variance} = \E_{\interest, \lambda} \mathrm{Var}(\zpivot).
		\end{align*}
		\item An estimate of the Kolmogorov-Smirnov distance between the one-tailed $p$-values obtained from $\Zpivot$ and the uniform distribution:
		\begin{align*}
			\mathrm{Max\ KS\ one\ tailed} = \max_{\interest, \lambda} \   \widehat{\mathrm{KS}}(\Phi(\zpivot), \mathrm{Unif}(0, 1)).
		\end{align*}
	\end{enumerate}

	We repeated this across 250 training runs, each with a different random initialisation, 500 epochs each of 100 steps.  This is one quarter the training used in each of the other examples presented in this paper, where a full training run is 1000 epochs of 200 steps.
			
	\section{Results}
	\label{section results}
	
	In the following sections we present the results from these experiments: NeuralCIs for the one-sample mean vs the $t$-test in Section \ref{section results t-test}, for Behrens-Fisher vs Welch in Section \ref{section results behrens fisher}, for partial biserial correlation vs profile likelihood ratio tests in Section \ref{section results biparcorr} and finally the toy problem measuring deviation from the maximal invariant in Section \ref{section results toy problem}.
	
	\subsection{$t$-Test}
	\label{section results t-test}

\begin{figure*}[tbp]
		\centering
		\includegraphics[width=\textwidth]{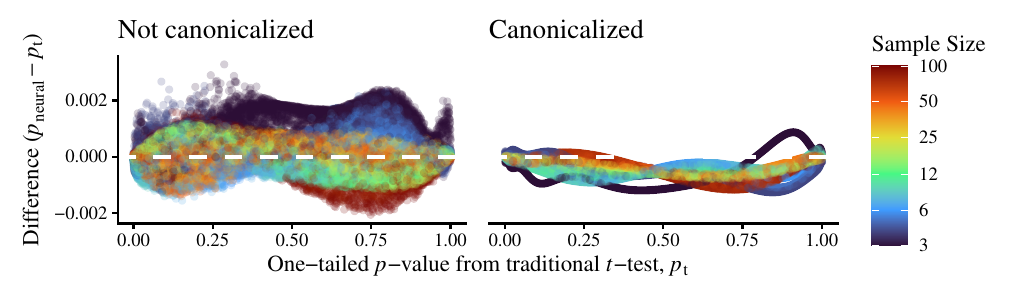}
		\caption{Comparison of NeuralCIs to traditional $t$-test.}
		\label{figure ttest}
	\end{figure*}

	In Figure \ref{figure ttest}, we scatter the difference between the NeuralCIs one-tailed $p$-value and that of the $t$-test, against that baseline $t$-test $p$.  Each plot is coloured according to sample size.  Both models produce one-tailed $p$-values close to those generated by the $t$-test.  For the non-canonicalized model, the absolute differences between the two were less than 0.0020 in 99.5\% of samples and less than 0.0033 in 100\% of samples; for the canonicalized model they were less than 0.0012 in 99.5\% and less than 0.0013 in 100\% of samples.
	
	\subsection{Behrens-Fisher}
	\label{section results behrens fisher}
	
\begin{figure*}[tbp]
		\centering
		\includegraphics[width=\textwidth]{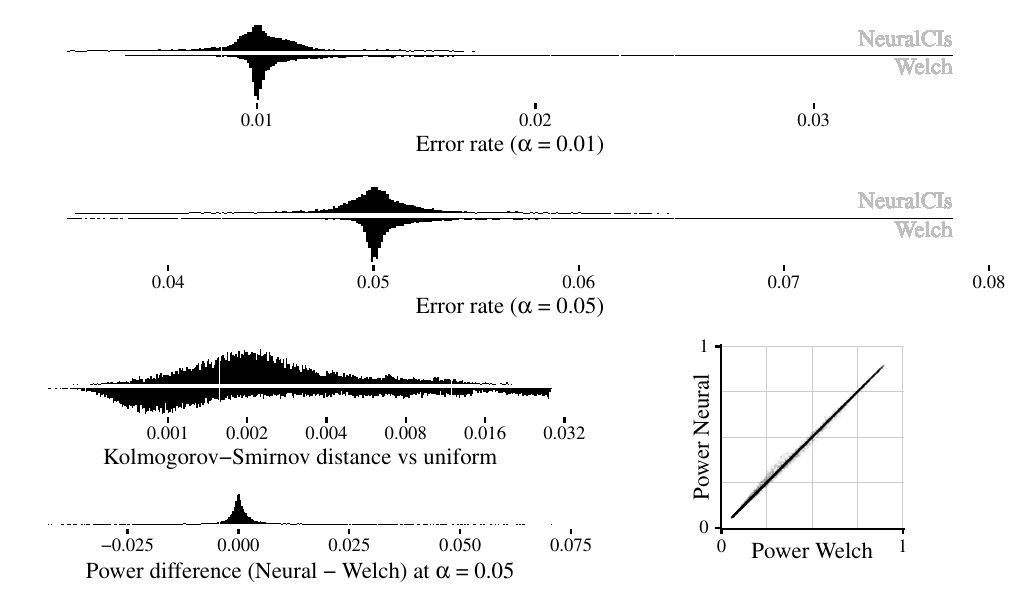}
		\caption{Comparison of NeuralCIs to traditional Welch test.}
		\label{welch figure}
	\end{figure*}

	Figure \ref{welch figure} compares false-positive rates and power between the NeuralCIs (top in each plot) and Welch (reflected below).  Although Welch shows a very slightly narrower peak, its size error in the worst cases is considerably worse than that of NeuralCIs.  The power difference between the two is negligible: less than one tenth of one percentage point in NeuralCIs' favour.
	
	\subsection{Partial biserial correlation}
	\label{section results biparcorr}
	
\begin{figure*}[tbp]
		\centering
		\includegraphics[width=\textwidth]{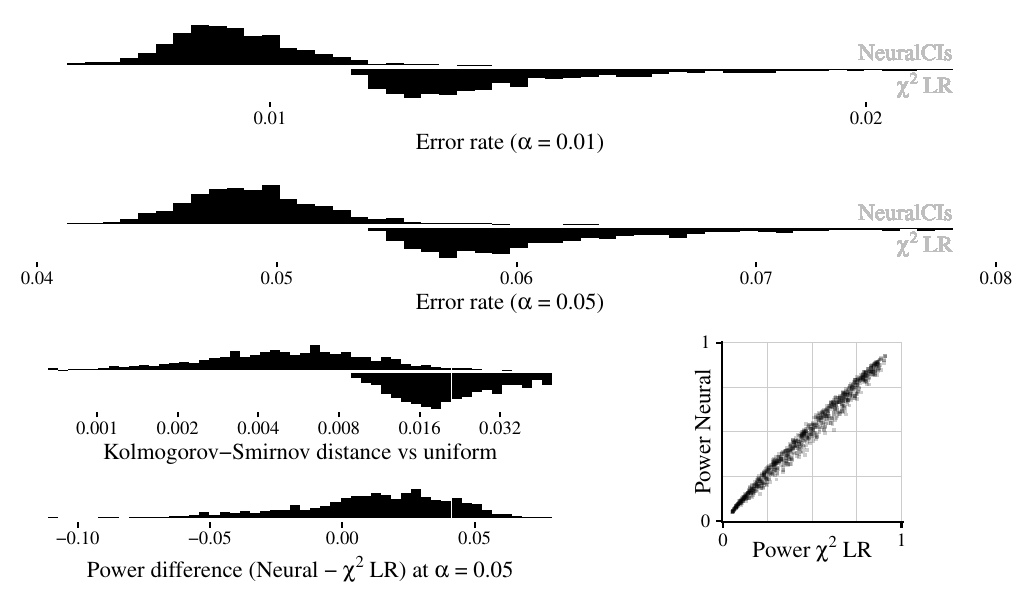}
		\caption{Comparison of NeuralCIs to traditional $\chi^2$ likelihood-ratio test in partial biserial correlation problem.  Power is size-adjusted, targeted at $\alpha = 0.05$.}
		\label{fig:biparcorr-chisq}
	\end{figure*}

	Results for the likelihood ratio $\chi^2$ test, presented in Figure \ref{fig:biparcorr-chisq}, show that, at these quite small sample sizes ($n \in [20, 100]$), the $\chi^2$ test is consistently rather liberal, where NeuralCIs is quite neatly clustered around the target rate.  Furthermore, NeuralCIs achieves on average 1.2 percentage points greater power. 
	
\begin{figure*}[tbp]
		\centering
		\includegraphics[width=\textwidth]{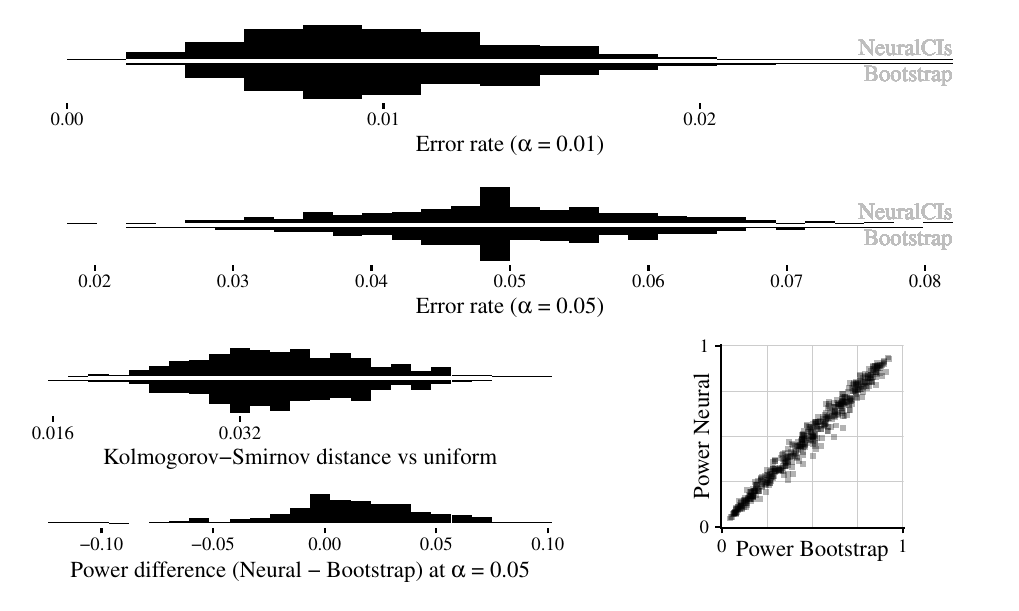}
		\caption{Comparison of NeuralCIs to the bootstrapped likelihood-ratio test in partial biserial correlation problem.  Power is size-adjusted, targeted at $\alpha = 0.05$.}
		\label{fig:biparcorr-bootlr}
	\end{figure*}

	 Results for the bootstrap likelihood ratio are presented in Figure \ref{fig:biparcorr-bootlr}.  Error rates are this time well centred around $\alpha$ for both tests (mean proportions below alpha are 0.010 and 0.049 for NeuralCIs and 0.010 and 0.050 for the bootstrap LR, at $\alpha = 0.01$ and $\alpha = 0.05$, respectively).  Again, NeuralCIs has a size-adjusted power advantage of 1.2 percentage points.
	
	The error rates for the bootstrap LR appear to be quite widely spread, but with only 500 $p$-values per parameter sample, we already expect these error rates to be widely spread simply by chance variation.  Table \ref{table lr sd comparison} shows that the spread (in terms of standard deviation) is close to the value expected from chance variation alone for both NeuralCIs and bootstrap LR; with only 500 $p$-values per parameter in the bootstrap comparison, modest variation in true test size would contribute very little to this spread\footnote{Note that the two would combine in terms of \textit{variance}, rather than standard deviation.} and therefore be hard to detect.  However, combining our knowledge from Figures \ref{fig:biparcorr-chisq} and \ref{fig:biparcorr-bootlr}, we know that NeuralCIs has reasonable control of error rates and a power advantage over bootstrap LR.  And importantly, it achieves this at speeds several orders of magnitude faster.  Generating a half a million $p$-values for this run, spread across null and power $p$-values, took a total of 39 hours even highly parallelized on a GPU; generating ten billion $p$-values (20,000 times as many) for our stand-alone neural simulation (Section \ref{section methodology neuralcis alone}) took only fifteen minutes.
	
	\begin{table}[tbp]
		\centering
		\begin{tabular}{lccccc}
			\multirow{2}{*}[-1.6ex]{\shortstack[l]{LR\\distribution}}& & \multicolumn{3}{c}{SD of error rate} & \multirow{2}{*}[-1.4ex]{\shortstack{Mean power\\difference}} \\
			\cmidrule(lr){3-5}
			
			& $\alpha$ & Expected & NeuralCIs & LR & \\
			\hline
			
			$\chi^2$ & 0.01 & 0.0001 & 0.0010 & 0.0022 & 0.011  \\ 
			& 0.05 & 0.0002 & 0.0028 & 0.0054 & 0.012  \\ 
			
			\hline
			Bootstrap & 0.01 & 0.0044 & 0.0043 & 0.0044 & 0.016  \\ 
			& 0.05 & 0.0097 & 0.0100 & 0.0099 & 0.012 \\
			\hline
			None (only & 0.01 & 0.0001 & 0.0011 & &  \\ 
			NeuralCIs) & 0.05 & 0.0002 & 0.0033 & & \\
			\hline
		\end{tabular} 
		\caption{NeuralCIs and profile LR approaches compared.  Expected standard deviation (SD) of error rates represents the standard deviation that would be expected, based on $\alpha$ and number $n_p$ of $p$-values per null parameter, if the error rate were perfectly calibrated to $\alpha$.  This is just the standard deviation of a binomial distribution divided by $n_p$, $\mathrm{Expected\ SD} = \sqrt{\frac{\alpha(1-\alpha)}{n_p}}$.  Mean power difference is NeuralCIs minus Profile LR.}
		\label{table lr sd comparison}
	\end{table}
	
\begin{figure*}[tbp]
		\centering
		\includegraphics[width=\textwidth]{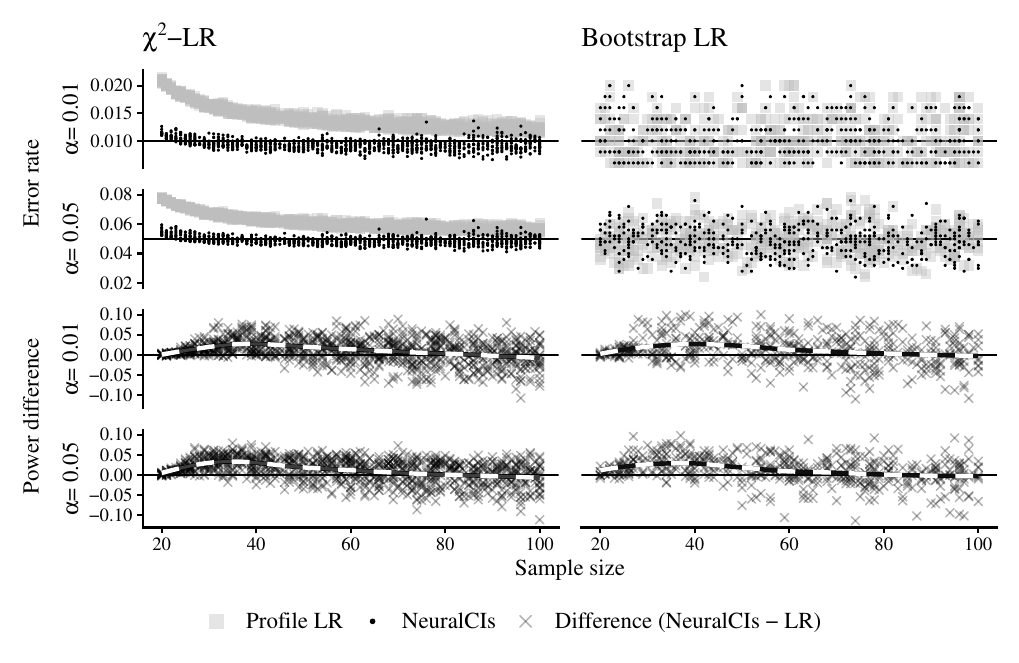}
		\caption{Results from Figures \ref{fig:biparcorr-chisq} and \ref{fig:biparcorr-bootlr}, broken down by sample size.  This must be interpreted with caution, as the parameter sampling distribution is different at each sample size.}
		\label{fig:biparcorr-n-comparison}
	\end{figure*}

	The comparison to the two profile LR approaches is much clearer when broken down by sample size.  Figure \ref{fig:biparcorr-n-comparison} shows that the advantages of NeuralCIs in terms of power and error rates are strongest in small to moderate sample sizes (sample sizes 20 to 70 in particular) and progressively narrowed or even reversed as the sample size approaches 100.  Between sample sizes 30-50, the power advantage appears to be around 3 percentage points on average.  We must be careful in interpreting these results, because the parameter sampling region in our test also grows with sample size\footnote{Furthermore, the boundary, to which extreme $H_1$ values of $\rho_{AB|C}$ are clipped is different at each $n$.}, but the overall pattern is consistent with expectations: profile LR is asymptotically optimal in terms of local power and the $\chi^2$ approximation is asymptotically correctly sized, but may be liberal and/or underpowered in finite samples.  The bootstrap can correct the size compared to the $\chi^2$ approximation, but with a very similar power disadvantage at smaller sample sizes (also to be expected, since the test statistic itself is the same).
	
\begin{figure*}[tbp]
		\centering
		\includegraphics[width=\textwidth]{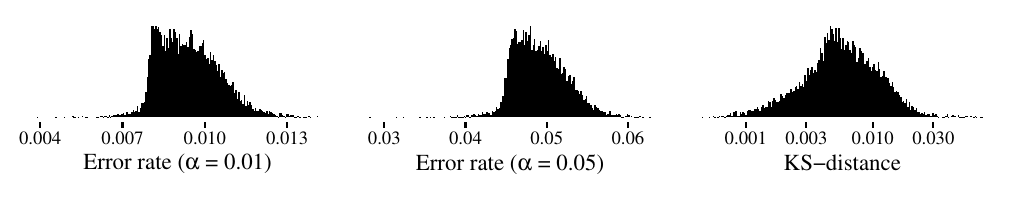}
		\caption{NeuralCIs false positive rates at $\alpha = 0.01$ and $\alpha = 0.05$ and Kolmogorov-Smirnov distances vs uniform for partial biserial correlation example.  Total of $10^4$ null hypothesis parameter samples plotted, each computed from $10^6$ datasets simulated at that null.}
		\label{fig:biparcorr-neural}
	\end{figure*}

	In Figure \ref{fig:biparcorr-neural}, we present false positive rates of NeuralCIs without comparison to another method.  NeuralCIs does not become degenerate at the boundaries in the same way as likelihood approaches, so these results can capture a broad range of possible null parameters.  The error rates are consistently quite close to the desired $\alpha$.  The ten billion $p$-values generated for Figure \ref{fig:biparcorr-neural} took fifteen minutes to run.
	
	In Section \ref{section convergence}, we present rates of non-convergence and boundary and degenerate samples in the two profile-LR approaches.  In the $\chi^2$ approximation, rates of non-convergence are mostly well below one in a thousand (or slightly above in a few cases).  Boundary samples are also almost entirely well below one in a thousand, except less than 1\% of parameter samples which were slightly above that\footnote{Most were even well below this: 81\% of main runs and 75\% of size-adjust runs had fewer than one in ten thousand boundary samples.} and degenerate samples (prevalence of 0\% or 100\%) were almost all below one in ten thousand.  For the bootstrap LR, we have far less ability to pre-plan avoiding boundary samples and most parameter samples have at least 1\% of their $p$-values based on non-convergence rates between 0.001 and 0.03, with the worst cases being occasionally as high as ten percent non-converged.  This applies only to the bootstrap distribution, since the same samples were used as the starting samples as in $\chi^2$.  
	
	So, while the convergence rates on the $\chi^2$ approximation look very comfortable, we urge a little caution in interpreting the bootstrap LR results.  Nonetheless, the pattern looks clean: both bootstrap LR and NeuralCIs are centred at the nominal $\alpha$, and the power advantage of NeuralCIs is consistent with that seen in the $\chi^2$ approach.
	
	\subsection{Deviation from the maximal invariant}
	\label{section results toy problem}

\begin{figure*}[tbp]
		\centering
		\includegraphics[width=\textwidth]{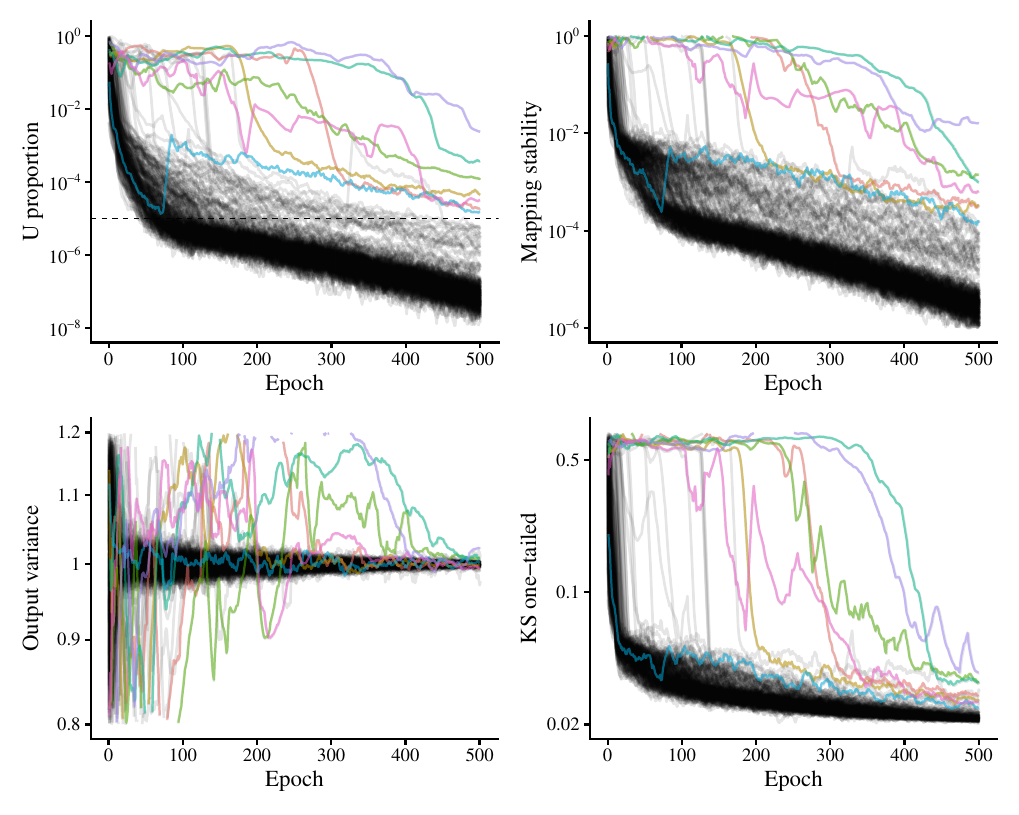}
		\caption{Is $\zpivot$ a function of the maximal invariant alone?  Each curve is smoothed using a uniform moving average with window size 10 epochs.  All $y$-axes are log-axes.  Those finishing with a smoothed $U$-proportion above $10^{-5}$ are coloured.}
		\label{fig:u-dependence}
	\end{figure*}

	Results of the $\uuu$-dependence toy problem are shown in Figure \ref{fig:u-dependence}.  The proportion of variance not accounted for by the maximal invariant (the ``$U$ proportion'') rapidly becomes negligible and continues to become smaller as training progresses, and the mapping becomes very stable.  All runs converge well to an output variance of 1 and a low KS-distance of $p$-values versus uniform.  Note that there is a floor to the KS-distances, imposed by the small grid-sample of $\pivot = e_1$; they cannot in this case be reduced below 0.02.
	
	Of the 250 random initializations, 7 converged noticeably less cleanly than the others, all having final smoothed $U$-proportion above $10^{-5}$ (coloured in Figure \ref{fig:u-dependence}).  While these higher final $U$-proportions are still effectively negligible, we trained these seven models for a further set of 500 epochs of 100 steps and present those results in Appendix \ref{section sine search further training}; all converge as expected after a further round of training.

	\section{Discussion}
	\label{section discussion}
	
	\subsection{Interpretation of results}
	
	The Behrens-Fisher problem, with continuous sufficient statistic of dimension $\nx = \nunknown$, represents a perfect example of the sort of problem NeuralCIs aims at.  Its performance is comparable to the Welch test, achieving considerably better worst-case false positive rates, with very similar power.  This is, of course, within pre-specified bounds on $\hat \sigma_2 / \hat \sigma_1$.
	
	On the other hand, our experiments with partial biserial correlation try to answer the question of whether this approach might be generalized to a cruder scenario, where ``ball-park estimates'' supplant sufficient statistics and even one of those estimates is not continuous.  In this particular case, NeuralCIs performs comparably with the bootstrap profile LR, but at a tiny fraction of the computational cost and with a small power advantage (1.2 percentage points on average), particularly at small sample sizes (a power advantage of around 3 percentage points for sample sizes from 30-50).  Notably, this was at sample sizes where the traditional Chi-squared approximation to the profile LR was quite inaccurate.  So, while the net must surely approximate (since one statistic is non-continuous), the approximation appears to be very effective.
	
	Generating $p$-value distributions close to the boundary was a significant problem in the profile LR framework, so much so that we had to abandon attempts to do so.  But in NeuralCIs, these can be generated quite readily and performed well.
	
	Finally, the $t$-test example shows how close NeuralCIs can come to discovering the classical pivot for a transformation model, while the toy problem (``deviation from the maximal invariant'') shows that this occurs consistently, even when other pivots do exist.
		
	\subsection{Limitations and future work}
	\label{section limitations and future work}
	
	We hope this to be the first of a line of steps in the NeuralCIs project, each step representing a current limitation of the project in its present form:
	
	\begin{enumerate}
		\item Fix the edge-of-$\innerTheta$ sampling issue highlighted in Appendix \ref{section ttest inner theta appendix}.
		
		\item The neural net dimensions have so far been chosen somewhat arbitrarily / in an \textit{ad hoc} fashion.  Optimizing the hyper-parameters to the net, such as number of layers, size of layers, number of training epochs, and learning rate schedule is an important next step.
		
		\item Invert our $p$-value net $\zpivot$ to train a further net that can generate confidence intervals for $\interest$.  See Section \ref{section confidence intervals}.
		
		\item Assess seed-to-seed variability in training.
		
		\item Train and evaluate several models relating to interesting core statistical problems (as with partial biserial correlation) and wrap these up in an R package for true amortization of the training process.
		
		\item Switch from explicitly differentiating and punishing Jacobians to training via the free-form flows (FFF) approach of \cite{draxler2024free}, to provide more scalable training and global diffeomorphism targeting (see Section \ref{section invertibility and tractability}).
		
		\item Currently only a single scalar interest parameter $\interest$ is supported.  We would like to extend this to capturing simultaneous confidence intervals and/or confidence regions on a vector-valued interest parameter.  One might, for example, make $\zpivot$ multidimensional, or model the combined distribution of several different NeuralCIs models, each trained on a different $\interest$.
		
		\item \label{relax nx}  The current architecture makes the severely limiting assumption that $\nx = \nunknown$.  When $\nx > \nunknown$, there may be multiple pivotal directions, some that are not informative about $\interest$.  The subtractive argument made in Section \ref{section power} was that the single remaining dimension after subtraction of nuisance parameters is expected to be highly informative; this no longer holds.  Expanding the framework to accept a higher-dimensional statistic would make it considerably more general.
		
		\item A longer term extension would be to replace $\zpivot$ and $\znuis$ with recurrent neural networks (RNNs)\footnote{Or, possibly, deep set architectures.}, that process individual i.i.d. observations $x_i$ sequentially while updating an internal state representing the accumulated evidence for or against the null.  This would depend on resolving point \ref{relax nx} above and would nonetheless require the fixed-dimensional learned state of the RNN to be able to hold sufficient information from an arbitrarily large sample.
	\end{enumerate}

	\subsubsection{Confidence intervals}
	\label{section confidence intervals}
	
	Generating confidence intervals from our current $p$-values approach will (we hope) be achieved by training a further net, using $\zpivot$ as a differentiable loss function.  The essence is to train a new net $\cinet(\xx, z)$ that maps statistics $\xx$, and a target $z \sim \mathcal N(0, 1)$, to a $\interest$ value that represents the boundary of a confidence interval.  Specifically, if we feed the $\interest$ value output by $\cinet$ back into $\zpivot$, the output from this,
	\begin{align*}
		z_\mathrm{CI}(\xx, z) &= \zpivot(\xx;\cinet(\xx, z)),
	\end{align*}
	should aim to equal the input $z$ value for all $z$ and plausible $\xx$.  In doing so, it inverts the $p$-value implied by $\zpivot$ into a confidence interval made up of the $\interest$ values at which $\xx$ would have the desired $p$-value.  By plausible $\xx$, we mean specifically $\xx \in \xset_{\innerTheta}$, see Section \ref{section param sampling}.  
	
	We can achieve this by repeatedly
	\begin{enumerate}
		\item sampling parameter $\uu \in \innerTheta$ using the denormalizing flow trained in Section \ref{section param sampling};
		\item sampling each $\xx \sim p_\xx(\cdot; \uu)$ from our simulator;
		\item randomly sampling $z \sim \Normal$ values;
		\item computing $z_{CI}(\xx, z) = \zpivot(\xx;\cinet(\xx, z))$;
		\item adjusting the weights in the $\interest_{CI}$ net to minimize some discrepancy metric $\Loss = \Delta(z_\mathrm{CI}, z)$ between these two.
	\end{enumerate}
		
	One may then obtain, say, a 95\% CI for given $\xx$ as
	\begin{align*}
		[\cinet(\xx, +1.96), \cinet(\xx, -1.96)],
	\end{align*}
	where the $\pm 1.96 = \Phi^{-1}((0.025, 0.975))$ values are the $\zpivot$ values corresponding to a two-tailed $p$-value of $0.05$.
		
	\section{Conclusion}
	\label{section conclusion}
	
	We have seen that a simple decomposed normalizing flow-style neural network structure can extract near-pivotal statistics along with their distributions and that they can perform inference on three classical problems to an accuracy level that is competitive with classical approaches, and even with some advantages.  The $t$-test is retrieved almost exactly, the Welch-test is bettered in terms of worst case error over a constrained variance-ratio range and the profile likelihood ratio for the partial biserial correlation is beaten in our simulation in terms of size-adjusted power.  
	
	The net can incorporate invariance information, making it more robust and giving it wider reach.  Furthermore, once the expense of fitting the net has been invested, the fitted net can then be used extremely cheaply thereafter (much cheaper than bootstraps or likelihood surface optimizers).  There are still considerable limitations, in particular the limitation that $\nx = \nunknown$ (if we are to have reasonable power).  Nevertheless, the work so far suggests that neural density methods based on normalizing flows could form a basis for a very general approach to frequentist inference.

	
	\backmatter
	
	\section*{Acknowledgements}
	
	This paper is dedicated to the memory of my mother, Marion, who died during the final stages of its preparation.
	
	\section*{Declarations}
	
	\bmhead{Use of generative AI}
	
	The core contribution of this paper (the decomposition of the normalizing flow presented in Section 3) and the accompanying codebase were initially developed independently by the author.  Anthropic's Claude and OpenAI's ChatGPT were used extensively throughout the theoretical development that followed, as a sounding board and as a critic: for brainstorming, for adversarial checking of the mathematical arguments, for directing the author towards relevant existing literature and theory (including, for instance, the classical theory of transformation models, which underpins the arguments in Appendix C), and for proofreading.  A few short utility functions in the simulation code were also drafted with AI assistance.  All arguments, experiments and conclusions have been verified by the author, who takes full responsibility for the content of this paper.
	
	\bmhead{Code availability}
	
	The code for this project is available at \url{github.com/philassheton/neuralcis}.
	
	\bmhead{Funding}
	
	This work was carried out by the author independently and received no external funding.
	
	\bmhead{Competing interests}
	
	The author declares no competing interests.

	 
	\bibliography{neuralcis.bib}

@Article{coccaro2020dnnlikelihood,
  author    = {Coccaro, Andrea and Pierini, Maurizio and Silvestrini, Luca and Torre, Riccardo},
  title     = {The {DNNLikelihood}: enhancing likelihood distribution with Deep Learning},
  journal   = {The European Physical Journal C},
  year      = {2020},
  volume    = {80},
  number    = {7},
  pages     = {664},
  month     = jul,
  issn      = {1434-6052},
  doi       = {10.1140/epjc/s10052-020-8230-1},
  publisher = {Springer Science and Business Media LLC},
}

@Article{welch1947generalization,
  author    = {Welch, Bernard L},
  title     = {The generalization of ‘Student's’ problem when several different population variances are involved},
  journal   = {Biometrika},
  year      = {1947},
  volume    = {34},
  number    = {1-2},
  pages     = {28--35},
  publisher = {Oxford University Press},
}

@Article{wilks1938large,
  author  = {Wilks, S. S.},
  title   = {The Large-Sample Distribution of the Likelihood Ratio for Testing Composite Hypotheses},
  journal = {The Annals of Mathematical Statistics},
  year    = {1938},
  volume  = {9},
  number  = {1},
  pages   = {60--62},
  doi     = {10.1214/aoms/1177732360},
}

@Article{papamakarios2021normalizing,
  author  = {Papamakarios, George and Nalisnick, Eric and Rezende, Danilo Jimenez and Mohamed, Shakir and Lakshminarayanan, Balaji},
  title   = {Normalizing flows for probabilistic modeling and inference},
  journal = {Journal of Machine Learning Research},
  year    = {2021},
  volume  = {22},
  number  = {57},
  pages   = {1--64},
}

@Article{olsson1982polyserial,
  author    = {Olsson, Ulf and Drasgow, Fritz and Dorans, Neil J},
  title     = {The polyserial correlation coefficient},
  journal   = {Psychometrika},
  year      = {1982},
  volume    = {47},
  number    = {3},
  pages     = {337--347},
  publisher = {Springer},
}

@Misc{cranmer2016approximating,
  author        = {Kyle Cranmer and Juan Pavez and Gilles Louppe},
  title         = {Approximating Likelihood Ratios with Calibrated Discriminative Classifiers},
  year          = {2015},
  archiveprefix = {arXiv},
  eprint        = {1506.02169},
  primaryclass  = {stat.AP},
  url           = {https://arxiv.org/abs/1506.02169},
}

@Article{dalmasso2024likelihood,
  author    = {Dalmasso, Niccol{\`o} and Masserano, Luca and Zhao, David and Izbicki, Rafael and Lee, Ann B},
  title     = {Likelihood-free frequentist inference: Bridging classical statistics and machine learning for reliable simulator-based inference},
  journal   = {Electronic Journal of Statistics},
  year      = {2024},
  volume    = {18},
  number    = {2},
  pages     = {5045--5090},
  publisher = {The Institute of Mathematical Statistics and the Bernoulli Society},
}

@Article{heinrich2022learning,
  author  = {Heinrich, Lukas},
  title   = {Learning optimal test statistics in the presence of nuisance parameters},
  journal = {arXiv preprint},
  year    = {2022},
  url     = {https://arxiv.org/abs/2203.13079},
}

@Article{zammit2025neural,
  author    = {Zammit-Mangion, Andrew and Sainsbury-Dale, Matthew and Huser, Rapha{\"e}l},
  title     = {Neural methods for amortized inference},
  journal   = {Annual Review of Statistics and Its Application},
  year      = {2025},
  volume    = {12},
  number    = {1},
  pages     = {311--335},
  publisher = {Annual Reviews},
}

@Book{davison1997bootstrap,
  title     = {Bootstrap methods and their application},
  publisher = {Cambridge university press},
  year      = {1997},
  author    = {Davison, Anthony Christopher and Hinkley, David Victor},
  number    = {1},
}

@Article{hall1988bootstrap,
  author    = {Hall, Peter and Martin, Michael A},
  title     = {On bootstrap resampling and iteration},
  journal   = {Biometrika},
  year      = {1988},
  volume    = {75},
  number    = {4},
  pages     = {661--671},
  publisher = {Oxford University Press},
}

@InProceedings{draxler2024free,
  author       = {Draxler, Felix and Sorrenson, Peter and Zimmermann, Lea and Rousselot, Armand and K{\"o}the, Ullrich},
  title        = {Free-form flows: Make any architecture a normalizing flow},
  booktitle    = {International Conference on Artificial Intelligence and Statistics},
  year         = {2024},
  pages        = {2197--2205},
  organization = {PMLR},
}

@Article{louppe2017learning,
  author  = {Louppe, Gilles and Kagan, Michael and Cranmer, Kyle},
  title   = {Learning to pivot with adversarial networks},
  journal = {Advances in neural information processing systems},
  year    = {2017},
  volume  = {30},
}

@Article{goodfellow2014generative,
  author  = {Goodfellow, Ian J and Pouget-Abadie, Jean and Mirza, Mehdi and Xu, Bing and Warde-Farley, David and Ozair, Sherjil and Courville, Aaron and Bengio, Yoshua},
  title   = {Generative adversarial nets},
  journal = {Advances in neural information processing systems},
  year    = {2014},
  volume  = {27},
}

@Article{al2024amortized,
  author    = {Al Kadhim, Ali and Prosper, Harrison B and Prosper, Olivia F},
  title     = {Amortized simulation-based frequentist inference for tractable and intractable likelihoods},
  journal   = {Machine Learning: Science and Technology},
  year      = {2024},
  volume    = {5},
  number    = {1},
  pages     = {015020},
  publisher = {IOP Publishing},
}

@Article{barndorff1988differential,
  author    = {Barndorff-Nielsen, OE and Jupp, Peter Edmund},
  title     = {Differential Geometry, Profile Likelihood, $ L $-Sufficiency and Composite Transformation Models},
  journal   = {The Annals of Statistics},
  year      = {1988},
  volume    = {16},
  number    = {3},
  pages     = {1009--1043},
  publisher = {Institute of Mathematical Statistics},
}

@InProceedings{rezende2015variational,
  author       = {Rezende, Danilo and Mohamed, Shakir},
  title        = {Variational inference with normalizing flows},
  booktitle    = {International conference on machine learning},
  year         = {2015},
  pages        = {1530--1538},
  organization = {PMLR},
}

@Article{paul2019review,
  author    = {Paul, Sudhir and Wang, You-Gan and Ullah, Insha},
  title     = {A review of the {Behrens--Fisher} problem and some of its analogs: Does the same size fit all?},
  journal   = {REVSTAT -- Statistical Journal},
  year      = {2019},
  volume    = {17},
  number    = {4},
  pages     = {563--597},
  doi       = {10.57805/revstat.v17i4.281},
  publisher = {Instituto Nacional de Estatistica},
}


	\appendix
	
	\section{Parameter sampling details}
	\label{parameter sampling appendix}
	
	The following section is a rather terse ``engineering decisions'' section, and not necessary to a rough understanding of the model.  Furthermore, we hope to improve the methodology below in future work.  For the reader wanting only to understand the key innovations in this paper, this section could reasonably be skipped.
	
	It should also be noted that this sampling scheme is designed for the ultimate goal to use NeuralCIs to generate confidence intervals, rather than directly $p$-values.  With a confidence interval, one starts at the sample, and explores compatible $\interest$ values.  On the other hand, with a $p$-value, one may choose a wildly inappropriate null $\interest_0$ for a given sample.  Our architecture does not currently employ effort to guard against such misadventure, since its goal is anyway a different destination.
	
	To define $\xtarget$ in a compact way, we define a bounding box $\boundingbox \subseteq \paramset$ over the parameter \textit{estimates} $\uuhat$ and define $\xtarget = \{\xx \in \xset | \uuhat(\xx) \in \boundingbox\}$.  In fact, we simplify each step by working in terms of $\uuhat$ (and which parameters produce $\uuhat$ that overlap) rather than $\xx$.
	
	To obtain a denormalizing flow (Section \ref{section importance sampling flows}) that can cheaply sample from $\outerTheta$ (as well as one that can sample cheaply from $\innerTheta$), a series of Metropolis-Hastings-style random walks, neural networks and denormalizing flows are fit to the sampling function:

	\begin{enumerate}
		\item \label{inner metropolis hastings step} Since $\uuhat$ should be an estimator of $\uu$, a number of MCMC chains are initialized starting at $\uu \in \boundingbox$.  At each Metropolis-Hastings-style step, a large number of samples $\uuhat(\xx)$ are drawn and an approximate bounding box, $\boundingbox_{\uuhat(\xx)}$ is computed as a Bonferroni-adjusted multiple of each estimate's respective standard deviation (Bonferroni-adjusted to capture 99.5\% of probability mass).  For $\uu$ where $\boundingbox_{\uuhat(\xx)} \cap \boundingbox \neq \varnothing$, the covariance matrix of the $\uuhat(\xx)$ is used to compute an approximate Fisher information, which will be used as the target distribution for the denormalizing flow in the following, while the proposal distribution for the following MCMC-jump is normal with the same covariance matrix.
		
		\item \label{inner feeler net step} A neural net is trained to model a surface that is a large negative number for inputs $\uu$ where $\boundingbox_{\uuhat(\xx)} \cap \boundingbox = \varnothing$ in step \ref{inner metropolis hastings step}, and to return the approximate square root Fisher information determinant from step \ref{inner metropolis hastings step} otherwise.
		
		\item \label{inner param sampling step} A denormalizing flow is fit to the surface from step \ref{inner feeler net step}.  This can be used to draw unlimited samples from $\innerTheta$.
		
		\item \label{is inside step} Samples $\uu \in \innerTheta$ are drawn from the denormalizing flow fit in step \ref{inner param sampling step}, and $\uuhat(\xx)$ values are sampled at each.  A neural net is trained to produce higher values where these $\uuhat$ values land and lower values where they do not.  This is done by assigning target 1 to the $\xx$ drawn from $\innerTheta$ and 0 to a set of samples drawn from a uniform distribution, and fitting a neural net with sum squared error\footnote{In later versions, we will experiment with cross-entropy error for this part of the net.}.  This net then represents $\xset_{\innerTheta}$ by all $\uuhat$ at which the net returns a value higher than\footnote{This value $\epsilon = 0.01$ we need to make dependent on the square root Fisher information modelled in Step \ref{inner feeler net step}; it is one of the two issues raised in Appendix \ref{section ttest inner theta appendix} that makes the parameter sampling ineffective for particularly wide-sampled $\uuhat$.} $\epsilon = 0.01$.
		
		\item Steps \ref{inner metropolis hastings step} to \ref{inner param sampling step} are repeated, except that generated $\uuhat$ are compared not  to $\boundingbox$ but instead to $\{\uuhat(\xx): \xx \in \xset_{\innerTheta}\}$, according to the net in step \ref{is inside step}.  The resulting denormalizing flow is a very cheap sampler from $\outerTheta$.
	\end{enumerate}

	In practice, maximum likelihood estimates $\uuhat$ can be rather expensive to compute.  We allow the use of other estimators $\uuhat(\xx)$.  Although these might not yield precisely the right volumes at each parameter, they should nonetheless scale in a ``close-enough'' way to provide ``good enough'' parameter sampling.  The one condition that is critical, if invariances are to be incorporated (Section \ref{section invariances}) is that the distribution of $\uuhat$ is also transported by the group action.
	
	\section{Experimental sampling for Partial biserial correlation}
	\label{appendix sampling biparcorr params}
	
	The following are the details of the parameter sampling scheme used in the evaluation of NeuralCIs in comparison with the two likelihood-ratio techniques.  Importantly, \textbf{this is not the parameter sampling scheme used for training} and it is rather narrow for smaller $n$.  The reason for the narrow selection of parameters is to avoid those parameters at which the likelihood optimisation (in the baseline likelihood ratio methods) will hit a large number of boundary values.
	
	The maximum and minimum values for each parameter are defined as follows.  Firstly, defining a target rate of hitting any given ``boundary'' as $\epsilon = 5 \times 10^{-5}$, we compute the maximum prevalence in $\Astar$ by inverting the probability that all values in $\Astar$ will equal 1:
	\begin{align*}
		\pi_{\max}(n) &= \epsilon^{1/n}, \\
		\pi_{\min}(n) &= 1 - \pi_{\max}(n).
	\end{align*}
	We further compute maximum and minimum values for the correlation $\rho_{BC}$ according to the Fisher $r$-to-$z$ transform:

	\begin{align}
		\rho_{\max}(n, r_{\max}) &= \tanh \left((z_{\max} - z_{\epsilon}) \sigma_{\mathrm{Fisher}}\right) \label{equation rho max}; \\
		z_{\max} &= \frac{\tanh^{-1}(r_{\max})}{\sigma_{\mathrm{Fisher}}}, \\
		z_{\epsilon} &= \Phi^{-1}(1 - \tfrac{\epsilon}{2}), \\
		\rho_{\min}(n, r_{\max}) &= -\rho_{\max}(n, r_{\max}), \\
		\sigma_{\mathrm{Fisher}} &
		= \frac{1}{\sqrt{n - 3}}.
	\end{align}
	where the division $\tfrac{\epsilon}{2}$ is to account for the fact that this is an approximation (even more so as it is used below).

	Meanwhile, the maximum value for the \textit{biserial} correlation $\rho_{\Astar C}$ is computed by treating $\rho_{\max}$ from (\ref{equation rho max}) as point-biserial and computing a conversion factor $k$ based on \cite{olsson1982polyserial}:
	\begin{align}
		\rho_{\max}^{\mathrm{biserial}}(n, \pi_{A^{\star}, r_{\max}}) &= k \rho_{\max}(n, r_{\max}/k),  \\
		k &= \frac{
			\sqrt{\pi_{\Astar}(1 - \pi_{\Astar})}
		}{
			\frac{
				e^{-0.5\Phi^{-1}(\pi_{\Astar})^2}
			}{
				\sqrt{2\pi}
			}
		}. \label{equation pointbi conv}
	\end{align}
	Finally, for the \textit{partial} biserial correlation $\rho_{\Astar B}$, we further subtract one from $n$:
	\begin{align*}
		\rho_{\max}^{\mathrm{partial\ biserial}}(n, \pi_{A^{\star}, r_{\max}}) &= \rho_{\max}^{\mathrm{biserial}}(n-1, \pi_{A^{\star}, r_{\max}}),
	\end{align*}
	with minimum values again set as the negative of the maximum.
	
	Since this sampling scheme tends anyway to bias samples towards the centre (particularly for low $n$), and also to cover a narrower region at lower $n$, we are satisfied to sample each parameter uniformly between its maximum and minimum values, including the sample size $n$.
	
	\subsection{Targeting power}
	
	As with Behrens-Fisher problem, we also sample separate $\rho_{AB|C}^{\mathrm{power}}$ values, which approximately target power ($1 - \beta$) uniformly sampled from $(0.05, 0.90)$, when used to assess the samples drawn under the null.  Equivalently, we draw $\beta \sim \mathrm{Unif}(0.10, 0.95)$ and:
	\begin{align}
		\rho_{AB|C}^{\mathrm{power}} = k \tanh(z_{\mathrm{power}} * \sigma_{\mathrm{Fisher}}); \\
		z_{\mathrm{power}} = z_{\mathrm{null}} \pm (z_{\alpha} + z_{\beta}) \label{equation z null}, \\
		z_{\mathrm{null}} = \frac{\tanh^{-1}(\rho_{AB|C}^{\mathrm{null}} / k)}{\sigma_{\mathrm{Fisher}}}, \\
		z_{\alpha} = \Phi^{-1} (1 - \alpha / 2), \\
		z_{\beta} = \Phi^{-1}(1 - \beta), \\
		\sigma_{\mathrm{Fisher}} = \frac{1}{\sqrt{n - 1 - 3}}.
	\end{align}
	where $k$ is computed as in (\ref{equation pointbi conv}) and the $\pm$ in (\ref{equation z null}) is randomized (except where only one of the two is within valid range).  Intuition note: dividing by $\sigma_{\mathrm{Fisher}}$ here transforms our $z$s from Fisher $r$-to-$z$ scale to ``standard-deviation-equals-one'' scale.  This makes the intermediate calculations much more intuitive.  Note also that in $\sigma_{\mathrm{Fisher}}$ we use $n-1-3$ rather than $n - 3$, because our correlation is partial.
	
	\section{Dimension-matched transformation models and the learned scalar coordinate $\zpivot$}
	\label{section transformation models}
	
	Let
	\begin{align*}
	\XX \in \RR^{\nunknown},
	\qquad
	\uu = (\interest, \llambda),
	\qquad
	\interest \in \RR,
	\qquad
	\llambda \in \RR^{\nunknown-1}.
	\end{align*}
	The normalizing flow learns the mapping
	\begin{align*}
		\XX \mapsto \ZZ = (\Zpivot, \Znuis),
	\end{align*}
	with
	\begin{align*}
	\Zpivot = \zpivot(\XX, \interest),
	\end{align*}
	so that the scalar coordinate has no direct nuisance input. The remaining coordinate
	\begin{align*}
	\Znuis = \znuis(\XX, \uu)
	\end{align*}
	may depend on the full parameter and absorbs the remaining $(\nunknown-1)$ directions.
	
	In this section, we examine the link in regular, full-rank, dimension-matched ($\nx = \nunknown$) transformation models, between $\zpivot$ and the profile likelihood ratio (LR).  In Section \ref{section maximal invariant}, we review the classical result that, in such transformation models, the profile likelihood ratio is a function of the maximal invariant.  Then in Section \ref{section zp to maxinv}, we argue that $\zpivot$ can reasonably be expected to land very close to this same maximal invariant, thereby closely linking $\zpivot$ to the profile likelihood ratio in such models.
	
	\subsection{Profile LR is a function of maximal invariant}
	\label{section maximal invariant}
	
	The link between the profile LR and the maximal invariant in a transformation model is well established; see, for example, \cite{barndorff1988differential}.  Here we present a simplified derivation, focussed on the particular case we have in mind for our architecture.  The core intuition is that transforming our data and parameters by $g$ multiplies both the numerator and denominator of the profile LR by the same Jacobian factor, which then cancels; so the profile LR itself is invariant under the group action and it must be a function of the maximal invariant.
	
	Assume the model is a transformation model. An $\nunknown$-dimensional Lie group $G$ acts smoothly on the data space and parameter space:
	\begin{align*}
	\xx \mapsto g\cdot\xx,
	\qquad
	\uu \mapsto g\cdot\uu.
	\end{align*}
	Assume the interest parameter is equivariant:
	\begin{align*}
	g \cdot \interest(\uu) \coloneqq \interest(g\cdot\uu).
	\end{align*}
	Thus $G$ acts jointly on pairs $(\xx, \interest)$ by
	\begin{align}
	\label{equation joint action}
	(\xx, \interest) \mapsto (g\xx, g\interest).
	\end{align}
	Assume the density is invariant up to the Jacobian factor $j_g(\xx)$:
	\begin{align*}
	p_{g\uu}(g\xx) = j_g(\xx) p_\uu(\xx),
	\end{align*}
	where $j_g(\xx)>0$ is the Jacobian factor and does not depend on $\uu$.
	Finally assume the joint action (\ref{equation joint action}) on $(\xx,\interest)$ is regular and full rank. Since
	\begin{align*}
	\dim(\xx, \interest) = \nunknown + 1
	\end{align*}
	and
	\begin{align*}
	\dim G = \nunknown,
	\end{align*}
	the quotient is locally one-dimensional. Let
	\begin{align*}
	m_\interest(\xx)
	\end{align*}
	denote a scalar maximal invariant of the joint action (\ref{equation joint action}).
	
	Define the profile likelihood
	\begin{align*}
	L_p(\interest;\xx)
	=
	\sup_{\uu: \interest(\uu) = \interest} p_\uu(\xx),
	\end{align*}
	and the profile likelihood ratio
	\begin{align*}
	\Lambda(\interest; \xx)
	=
	\frac{L_p(\interest; \xx)}
	{\sup_\uu p_\uu(\xx)}.
	\end{align*}
	Then
	\begin{align*}
		L_p(g\interest; g\xx)
		&=
		\sup_{\uu': \interest(\uu') = g\interest}
		p_{\uu'} (g\xx)\\
		&=
		\sup_{\uu: \interest(\uu) = \interest}
		p_{g\uu} (g\xx)\\
		&=
		\sup_{\uu: \interest(\uu) = \interest}
		j_g(\xx) p_\uu(\xx)\\
		&=
		j_g(\xx) L_p(\interest; \xx).
	\end{align*}
	Similarly,
	\begin{align*}
	\sup_{\uu'} p_{\uu'}(g\xx)
	=
	j_g(\xx)\sup_\uu p_\uu(\xx).
	\end{align*}
	(That is, for fixed $g$, we find the supremum across all $\uu$.)
	
	Therefore the Jacobian factor cancels in the likelihood ratio:
	\begin{align*}
	\Lambda(g\interest; g\xx)
	=
	\Lambda(\interest; \xx).
	\end{align*}
	
	So $\Lambda$ is invariant under the joint group action on $(\xx,\interest)$. Since $m_\interest(\xx)$ is maximal invariant,
	\begin{align*}
	\Lambda(\interest; \xx) = h(m_\interest(\xx)),
	\end{align*}
	for some scalar function $h$.

	\subsubsection{Gaussian location and other examples}
	\label{section gaussian location}
	
	\newcommand{\SSigma}{\mathbf{\Sigma}}
	\newcommand{\aaa}{\mathbf{a}}
	\newcommand{\xpsi}{x_{\interest}}
	\newcommand{\xlambda}{\xx_{\llambda}}
	\newcommand{\Xpsi}{X_{\interest}}
	\newcommand{\Xlambda}{\XX_{\llambda}}
	\newcommand{\apsi}{a_{\interest}}
	\newcommand{\alambda}{\mathbf{a}_{\llambda}}
	\newcommand{\dlambda}{\mathbf{d}_{\llambda}}
	
	For the simple Gaussian location model, $\XX \sim \Normal(\uu, \SSigma)$, with fixed covariance matrix $\SSigma = \iinfo^{-1}$, the unrestricted MLE of $\uu$ is $\uuhat = \XX$:
	\begin{align*}
		\XX = \uuhat =  \mat{\interesthat\\\lambdahat} \sim \Normal\left(\mat{\interest\\\llambda}, \mat{\info_{\interest\interest} & \iinfo_{\interest\llambda} \\ \iinfo_{\llambda\interest} & \iinfo_{\llambda\llambda}}^{-1}\right).
	\end{align*}
	The group action is a translation
	\begin{align*}
		\mat{\interesthat \\ \lambdahat \\ \interest} \mapsto \mat{\interesthat + \apsi \\ \lambdahat + \alambda \\ \interest + \apsi},
	\end{align*}
	with a simple maximal invariant
	\begin{align*}
		\pivot = \interesthat - \interest
	\end{align*}
	and quadratic log likelihood surface
	\begin{align*}
		\ell(\uu; \uuhat) - \ellhat = -\tfrac{1}{2}(\uu - \uuhat)^\top \iinfo (\uu - \uuhat),
	\end{align*}
	where $\ellhat = \ell(\uuhat; \uuhat)$ is a constant (independent of $\uu$) and represents the central (peak) log likelihood value at $\uu = \uuhat$.
	
	Writing also $\dlambda = \lambdahat - \llambda$, then the profile log likelihood $\ell_p$, and thereby the log likelihood ratio $\log \Lambda(\interest)$, is obtained by a simple optimization across this quadratic,
	\begin{align*}
		-2 \log \Lambda(\interest) = \phantom{aaa.}& \\
		-2\left(\ell_p(\interest; \interesthat) -  \ellhat_p\right)
		&= \min_{\llambda} \  (\uu - \uuhat)^\top \iinfo (\uu - \uuhat) \\
		&= \min_{\llambda} \ 
		\mat{\pivot\\\dlambda}^\top 
		\mat{\info_{\interest\interest} &
			\iinfo_{\interest\llambda} \\ \iinfo_{\llambda\interest} & \iinfo_{\llambda\llambda}}
		\mat{\pivot\\\dlambda} \\
		&= \min_{\dlambda}\  \left[
		\info_{\interest\interest} \pivot^2
		+ 2\pivot \iinfo_{\interest\llambda}\dlambda
		+ \dlambda^\top \iinfo_{\llambda\llambda} \dlambda
		\right] \\
		&= \pivot^2\left(
		\info_{\interest\interest}
		- \iinfo_{\interest\llambda} \iinfo_{\llambda\llambda}^{-1} \iinfo_{\llambda\interest}
		\right) \\
		&= \pivot^2 \effinfo \\
		&= (\interest - \interesthat)^2 \effinfo
	\end{align*}
	where $\ellhat_p = \ell_p(\interesthat;\interesthat)$ is again a constant and $\effinfo$ is the efficient information, which is constant across this quadratic log likelihood function.
	
	This simple transformation model illustrates the profile log likelihood ratio as a function of the maximal invariant $\pivot$ of $(\uuhat = \xx, \interest)$.

	\subsection{Why we think $\zpivot$ will tend towards $\pivot$}
	\label{section zp to maxinv}
	
	A Gaussianization of the maximal invariant $\pivot$ is a very natural target for $\zpivot$.  Define $G_n$ to be the $(\nunknown-1)$-dimensional nuisance subgroup of $G$ that leaves $\interest$ stationary.  Then $\pivot$ defines a foliation of $\xset$ such that probability mass is only moved by $G_n$ within each given leaf, and never between.  The foliation provides the perfect structure to model with $\znuis$, leaving the distribution across the leaves $p(\pivot)$, which is a pivot independent of $\llambda$.
	
	In this section we will  consider whether there are other viable targets for $\zpivot$.  Transform $\XX$ at fixed $\interest = \interest_0$ as
	\begin{align*}
		\XX_{(\interest_0, \llambda)} \leftrightarrow (M, \UU_{\llambda}),
		\qquad
		M = {\pivot}_0(\XX_{(\interest_0, \llambda)}),
		\qquad
		\UU_{\llambda} \in \RR^{\nx - 1},
	\end{align*}
	and correspondingly overload
	\begin{align*}
		\zpivot(\xx; \interest_0) \leftrightarrow
		\zpivot(m, \uuu); \qquad \xx \leftrightarrow (m, \uuu).	
	\end{align*} 
	We want to know if a $\zpivot(M, \UU_{\llambda})$ that depends not only on $M$, but also on $\UU_{\llambda}$ can still be standard normal for all $\llambda$.
	
	The short answer is: yes, there are sometimes loss-optimal forms of $\zpivot$ that depend not just on $m$ but also on $\uuu$, but all require considerable coordination of cancellations across $\uuu$-space that is unlikely to arise in any substantial form from the combination of a random initialisation and stochastic gradient descent optimisation.
	
	In Section \ref{section u dependence marginalized}, we start by assuming that $\zpivot(m, \uuu)$ is monotone increasing in $m$.  In this case, any dependence on $\uuu$ must be removed in expectation as $\llambda$ slides $\UU_{\llambda}$ around space, with any $\uuu$-dependence lost at the back end of the distribution, counterbalanced by a coordinated opposite dependence at the front end of the $\llambda$-induced movement of the probability mass.
	
	We give a simple two-dimensional example (that is, with scalar $U_{\lambda}$): $M \bot U_{\lambda}$, $U_\lambda \sim \mathrm{Unif}(\lambda, \lambda + 1)$.  In this case, a dependence on $U_{\lambda}$ that is periodic (with period 1) will average out to a constant across the uniform $U_{\lambda}$ at all $\lambda$.  This clearly shows the cancellation concept: as the uniform distribution is moved along, the portion of $\zpivot$ that is no longer supported is replaced by exactly the same shape in a new region of $u$ that previously had not been supported.  The loss is blind to such a specific, coordinated $u$-dependence and will neither encourage nor discourage it: without the appearance of such a coherent, coordinated pattern in the random initialization, it can only be arrived at during training by random drift in that direction.  But the component of any random drift in such an extremely specific direction could be expected to be negligible, as could the probability of it appearing in any significant proportion in the random initialization in the first place.
	
	In Section \ref{section folding in m} we then explore the possibility that $\zpivot$ is non-monotone in $m$.  This would make it possible for pairs of local $\uuu$-dependencies to cancel each other, without needing the same level of global coordination.  However, for $\zpivot$ to be non-monotone in $m$,  satisfying our architectural constraint $\tfrac{\partial \zpivot}{\partial \interest} < 0$ in (\ref{equation constrain negative zp by psi}) while the full map still remains a diffeomorphism, itself requires similarly complex, large-scale coordinated patterns.  These are, similarly, unlikely to arise from our random initialization and training.
	
	So, in summary, while it is technically possible for $\zpivot$ to stray far from the maximal invariant $\pivot$, it would require accidental construction of such high levels of coordination in the random initialization, that we consider it highly unlikely, except perhaps for a very small component.  Under the simple independent-noise benchmark in Section \ref{section maxinv power}, a small contamination like this should have a very mild effect on power (only second-order in the contamination amplitude).

	\subsubsection{Dependence on $\UU$ that disappears after marginalization}
	\label{section u dependence marginalized}
	
	Assume that $\zpivot(m, \uuu)$ is strictly increasing in $m$ for each fixed $\uuu$.  Define the inverse of $\zpivot$ at each $\uuu$ as $\mboundary(\uuu)$ according to
	\begin{align*}
		\zpivot(\mboundary(\uuu), \uuu) = z.
	\end{align*}
	We know that one excellent target for $\zpivot(M, \UU_{\llambda})$ would be a Gaussianization of $M$, call it $\zpivot^*$.  In this case the inverse is independent of $\uuu$, call it $\mboundaryNoU$:
	\begin{align*}
		\zpivot^*(\mboundaryNoU, \uuu) = z,
	\end{align*}
	for all $\uuu$ and $z$.
	
	In order that $\zpivot$ be pivotal, we need that $\zpivot(M, \UU_{\llambda})$ have the same distribution as $\zpivot^*(M, \UU_{\llambda})$ for all $\llambda$.  If we think of $\mboundaryNoU$ for any fixed $z$ as a boundary that is straight in $\uuu$, and $\mboundary$ as a boundary that is wiggly in $\uuu$, they must cross each other so that $\mboundary$ can exchange increased probability mass in one region of $\uuu$ for decreased probability mass in another.  We can represent these regions of gained or lost probability mass as regions of $\pm 1$ (with zero everywhere else),
	\begin{align*}
		\mswap(m, \uuu) = \mathbf{1}\{m < \mboundary(\uuu)\}
		- \mathbf{1}\{m < \mboundaryNoU\},
	\end{align*}
	and pivotality requires that the same amount of probability is always exchanged between positive and negative regions,
	\begin{align*}
		\E[\mswap(M, \UU_{\llambda})] = 0,
	\end{align*}
	\textbf{for all $z$ and $\llambda$.}
	
	The final bolded part, that this must hold for all $z$ and $\llambda$ is the critical part.  Since our group acts in a full-rank way on the data, $\UU_{\llambda}$ is constantly moved around space by $\llambda$, and certain regions gain probability mass, while others lose it.  These two must be constantly offset and the result is that any dependence on $\uuu$ (if a feasible one should exist at all) must be highly globally structured across space.
	
	This argument does not depend on the scale of the $m$-component, since $m$ is defined only up to a monotone transformation.  It may be made arbitrarily weak while still requiring the same cancellation in $\uuu$-components as $\llambda$ moves probability mass around space.  In the limiting case where $\zpivot$ becomes independent of $m$ altogether, the $\llambda$-induced motion of $\UU_{\llambda}$ in all directions would demand the same globally coordinated cancellations, to maintain pivotality at all $\llambda$, if such a construction can be made pivotal at all.
	
	The argument is most easily visualized with an example.  Suppose that $\UU_{\llambda} = U_\lambda$ is scalar, $M \bot U$ under each scalar $\lambda$ and
	\begin{align*}
		U_\lambda \sim \mathrm{Unif}(\lambda, \lambda + 1).
	\end{align*}
	If we choose a wiggly boundary $\mboundary$ such that, when fed through the CDF $F_M$ of $M$, it generates a sine wave in $u$,
	\begin{align*}
		F_M(\mboundary(u)) = F_M(\mboundaryNoU) + a_z \sin (2\pi nu),
	\end{align*}
	at any integer frequency $n \in \mathbb{Z}^+$ and any set of amplitudes $a_z$, such that the right-hand side remains in $[0, 1]$ and increasing in $z$, then
	\begin{align*}
		P(\zpivot(M, U_\lambda) < z) &= \E[F_M(\mboundary(U_\lambda))] \\ 
		&= F_M(\mboundaryNoU) + a_z \E[\sin(2\pi n U_\lambda)] \\ 
		&= F_M(\mboundaryNoU).
	\end{align*}
	
	This example is a very clean illustration of the broader point: as the uniform distribution is translated along $u$ by $\lambda$, the wiggle that it leaves behind at the back must be compensated by another picked up at the front.  In this case, that means that the dependence on $\uuu$ must be periodic (with period 1) across the whole of $\uuu$.
	
	As the example shows, such solutions may exist, but they also require global coordination of the dependence on $\uuu$.  If some such globally coordinated dependence is already present in the random initialisation, the KL loss may have no incentive to remove it. However, this is highly unlikely to form more than a very small part of the initial $\uuu$-dependence. Most randomly initialized $\uuu$-dependencies will fail to produce the required cancellations for all $z$ and $\llambda$, and will therefore be ironed out during training.  
	
	The loss itself is ambivalent to such structures, so training may allow a random walk in that direction, but we could expect in most cases that the component of any random drift that is directed so specifically in that direction should be negligible.  Furthermore, against the ambivalence of the loss, the architecture itself may well prefer a simpler solution (flat in $\uuu$).
	
	\subsubsection{Folding in $M$}
	\label{section folding in m}
	
	If $\zpivot(m,\uuu;\interest)$ is not monotone in $m$, then, for fixed $\uuu$, the set
	\begin{align*}
		\left\{m : \zpivot(m, \uuu; \interest) \leq z\right\}
	\end{align*}
	may contain several disjoint intervals. Probability gained on one branch by a detour of the $\zpivot$-contour through $\uuu$ can then be offset by probability lost on another, while $\znuis$ retains the information needed to keep the full transformation invertible.
	
	Such a fold cannot stand alone. Since the full map
	\begin{align*}
		(m,\uuu) \mapsto (\zpivot,\znuis)
	\end{align*}
	is a diffeomorphism, the contours of $\zpivot$ are smooth, non-intersecting surfaces: they cannot terminate or cross, but rather form continuous ``snakes'' in space, enclosing volumes that encode how the model is distributed.  A turning point $\tfrac{\partial \zpivot}{\partial m} = 0$ can therefore only occur where the contour runs out through $\uuu$, turns back and returns at a different value of $m$.  A single fold has a ripple effect, requiring a coordinated family of ``snaking'' contours over at least a substantial neighbourhood, to maintain a diffeomorphism that still captures the model densities.
	
	Furthermore, our architectural constraint 
	\begin{align*}
		\frac{\partial \zpivot(\xx;\interest)}{\partial \interest} < 0
	\end{align*}
	from (\ref{equation constrain negative zp by psi}) is evaluated at fixed $\xx$: increasing $\interest$ moves every point downward in $\zpivot$.  Write $\Zpivot = \zpivot(\XX; \interest)$.   Then calibration $\Zpivot \sim \Normal$ must be preserved, while the data distribution itself moves under the infinitesimal interest action $\VV_\interest = \vv_\interest(\XX;\interest, \llambda)$.  Along this movement, we require that the flux (denoted $A$ below) of probability density into and out of a given $\Zpivot$ level set, as $\interest$ is tweaked, must be tracked/counterbalanced by a shift in the value of $\Zpivot$ (denoted $B$):
	\begin{align}
		\nonumber
		\frac{\dd}{\dd \interest} P_{\interest, \llambda}(\Zpivot \leq z) 
		&= -\phi(z)\E_{\interest, \llambda}\left[
		\frac{\mathrm{d}}{\mathrm{d}\interest} \Zpivot
		\middle| \Zpivot = z
		\right] \\
		\label{equation terms A and B}
		&= -\phi(z) \E_{\interest, \llambda}\left[
		\smash[b]{\underbrace{
				\vphantom{\frac{\partial DUMMY}{\partial \interest}}
				\nabla_\xx \Zpivot ^\top \VV_\interest
				\rule[-3ex]{0pt}{3ex}
			}_{A}}
		+
		\smash[b]{\underbrace{
				\frac{\partial \Zpivot}{\partial \interest}
				\rule[-3ex]{0pt}{3ex}
			}_{B}}		
		\middle| \Zpivot = z
		\right]
		= 0
	\end{align}
	for all $z$, $\llambda$ and $\interest$.
	
	Because the interest transformation is not entirely contained within the nuisance orbits, its motion $\VV_\interest$ has a non-zero component in the $m$-direction; we choose the orientation of $m$ so that
	\begin{align*}
		\nabla_\xx m ^\top \vv_\interest > 0.
	\end{align*}
	In the case where $\zpivot$ is strictly increasing in $m$, there can be a natural pointwise cancellation between terms $A$ and $B$ in (\ref{equation terms A and B}).  On the other hand, wherever $\zpivot$ is decreasing in $m$, the $m$-component of term $A$ reinforces term $B$, which must then be cancelled by motion in $u$ or by coordination with other $m$ that are mapped to the same $\zpivot$.
	
	A folded solution is therefore not impossible and it can in principle hide $\uuu$-dependencies from our loss, since gains and losses associated with different $m$ sharing the same $\zpivot$ can be cancelled.  But it must coordinate the geometry of an entire family of contours with compensating motion in $\uuu$ or on other branches, consistently across $z$, $\llambda$ and $\interest$. This is then (as with Section \ref{section u dependence marginalized}) a highly organized construction, rather than a local perturbation, that random initialization and gradient training would naturally struggle to produce.
	
	\subsubsection{Consequence of contamination for power}
	\label{section maxinv power}
	
	Our conclusion is that there may be a very small $\uuu$-dependent component in $\zpivot$, which may be rather uninformative about $\interest$; indeed it may be pure noise.  But there is unlikely to be a larger one.  To illustrate the likely scale of this effect, we consider below the simple benchmark in which the contaminating component is independent standard normal noise carrying no $\interest$-signal.  We show that noncentrality in this case is attenuated only quadratically with the amplitude of the mixed-in noise.
	
	First, consider the canonical $\Zpivot^*$, which depends only on $M$.  Suppose that under the alternative hypothesis, $\Zpivot^*$ is translated by $\delta_\interest$:
	\begin{align*}
		\Zpivot^* \sim \Normal(\delta_\interest, 1),
	\end{align*}
	so that $\delta_\interest$ is the noncentrality of $\Zpivot^*$.
	
	Suppose we mix in some $\UU$-derived standard normal noise $W \sim \Normal(0, 1)$, such that $W \bot \Zpivot^*$ and $W$ carries no $\interest$-signal at all, then
	\begin{align*}
		\Zpivot = \frac{\Zpivot^* + \epsilon W}{\sqrt{1 + \epsilon^2}} \sim \Normal\left(\frac{\delta_\interest}{\sqrt{1 + \epsilon^2}}, 1\right). 
	\end{align*}
	So the non-centrality of $\Zpivot$ itself is divided by the denominator $\sqrt{1 + \epsilon^2}$.  Or, expanding around $\epsilon=0$, its non-centrality is \textbf{multiplied} by
	\begin{align*}
		1 - \tfrac{1}{2}\epsilon^2 + \mathcal O(\epsilon^4).
	\end{align*}
	
	So the loss of noncentrality, and hence the resulting loss of power (which is a smooth function thereof), is only second-order in $\epsilon$.	
	
	\section{Further results}
	
	Here we present further results to support and further develop those presented in the main sections.
	
	\subsection{$t$-Test: the edges of $\innerTheta$}
	\label{section ttest inner theta appendix}

\begin{figure*}[tbp]
		\centering
		\includegraphics[width=\textwidth]{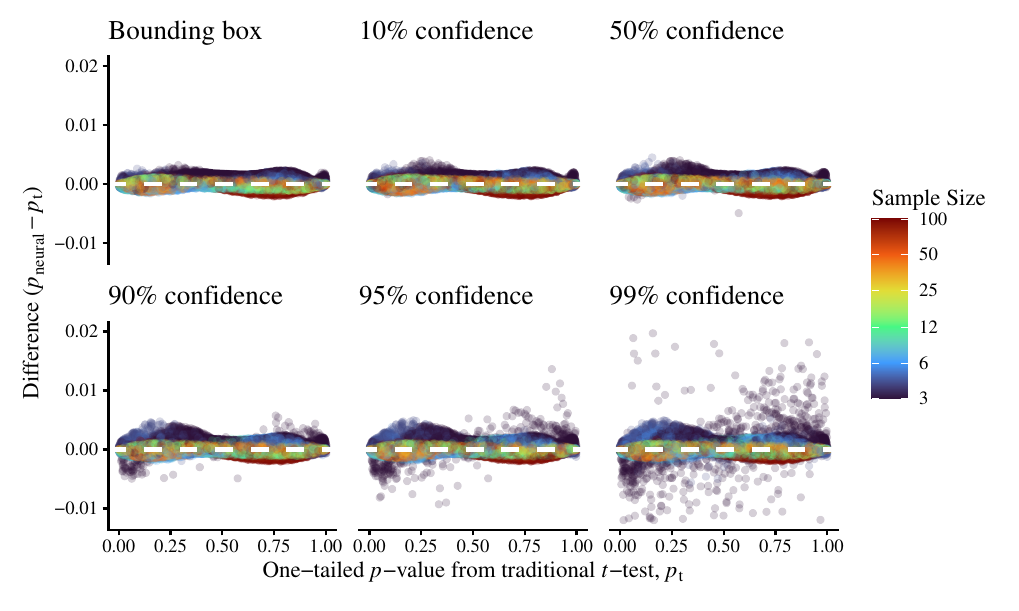}
		\caption{Comparison of Figure \ref{figure ttest} (parameters drawn from the estimates bounding box) with results obtained when extending the parameter sampling distribution into the edges of $\innerTheta$.  Other plots draw parameters from a union of Bonferroni confidence regions computed around the boundary points of the bounding box; percentages in individual plot titles represent the confidence levels of these confidence regions.}
		\label{fig:ttest-bonferroni}
	\end{figure*}

	In Figure \ref{fig:ttest-bonferroni}, we compare the results for the $t$-test model with no canonicalization, across a number of different parameter testing regions.  Top left reproduces Figure \ref{figure ttest} from the main text, where \textit{parameters} were drawn only from the estimates bounding box $\boundingbox$.  Of course, our goal is to generate valid $p$-values for all $\uu \in \innerTheta \supset \boundingbox$.
	
	To test performance outside of the bounding box, we draw a Bonferroni-adjusted parameter sampling quadrilateral at each $n$, that extends beyond the estimates box.  At a given $n$, consider the two-dimensional space $(\mu, \sigma)$, on which the estimates box is drawn.  At each point in this space, we compute a separate $p$-value in each dimension ($\mu$ and $\sigma$), taking the $(\mu, \sigma)$ coordinates at that point as the respective nulls, and the closest point on the bounding box as the estimates $(m, s) = (\hat \mu, \hat \sigma)$.   These $\nunknown = 2$ $p$-values are then combined by Bonferroni.  Those with a $p$-value greater than one-minus-confidence-level are allowed for inclusion in the plot, and the first 50000 in each case are retained for plotting.
	
	It is striking that the larger sample sizes perform relatively well at all confidence levels, where the small $n$ cases perform progressively worse towards the edge of the box.  Our current hypothesis is that the current sampling scheme has two weak-points with regard to parameters with wider spread estimates (as is the case with low $n$ in the $t$-test example):
	
	\begin{enumerate}
		\item \label{is inside step is bad} Step \ref{is inside step} in Appendix \ref{parameter sampling appendix}, which models the distribution of samples that can be drawn from $\innerTheta$, is based effectively on the absolute height of the probability density, which is much lower for a parameter with wider spread estimates (e.g. large $\sigma$ at small $n$ in the $t$-test).  This leads to such parameters being sampled less broadly than those with a tighter sampling distribution.  We might remedy this by making use of the net trained in step \ref{inner feeler net step}, which encodes the inverse of the volume occupied by the estimates and is a proxy for the height of the density.
		
		\item \label{jeffreys bad} The current sampling scheme, based on the Jeffreys prior, only takes account of the volume of the sampling distribution at that parameter, but does \textit{not} take account of the relative volume of space that it can fill.  If either of $\boundingbox$ or $\innerTheta$ is large compared to the spread of the sampling distribution of the estimates, our parameter sampling will overweight parameters with more tightly spread estimates, potentially catastrophically.  We might remedy this by weighting each parameter relative to probability of collision of the sampling distribution with $\boundingbox$ or $\xset_{\innerTheta}$, rather than (as at present) the probability of the collision of the sampling distribution with a single point.
	\end{enumerate}
	
	Point \ref{jeffreys bad} is more easily understood with a concrete example.  In this $t$-test example, we \textit{do} need a higher probability of drawing small $\sigma$ values, since their sampling distributions reach less far through space, and so a smaller volume of them overlap the bounding box; with equal sampling probability, they would then be underweight.  
	
	If the region we were aiming to overlap with were a single point, the Jeffreys prior would correct for this exactly: we would pull the same proportion of small $\sigma$ parameters as large $\sigma$.  But as we extend the bounding box outwards, these up-weighted small-$\sigma$ values start to dominate, as the same amount of space \textit{inside} the bounding box can accommodate far more small-$\sigma$ parameter samples than large-$\sigma$ ones.  This disproportionately impacts small sample sizes, because, particularly for $n=3$, the sample estimate $s$ is quite noisy and so a comparatively huge range of $\sigma$ values must be modelled.  
	
	We are developing an improved sampling scheme, that takes account of the size of the estimates box $\boundingbox$ when weighting $\innerTheta$, and of the size of $\innerTheta$ when weighting $\outerTheta$, as well as attempting to address point \ref{is inside step is bad} through the use of the modelled Cholesky determinants.
	
	\subsection{Non-converged, boundary and degenerate cases (partial biserial correlation)}
	\label{section convergence}

\begin{figure*}[tbp]
		\centering
		\includegraphics[width=\textwidth]{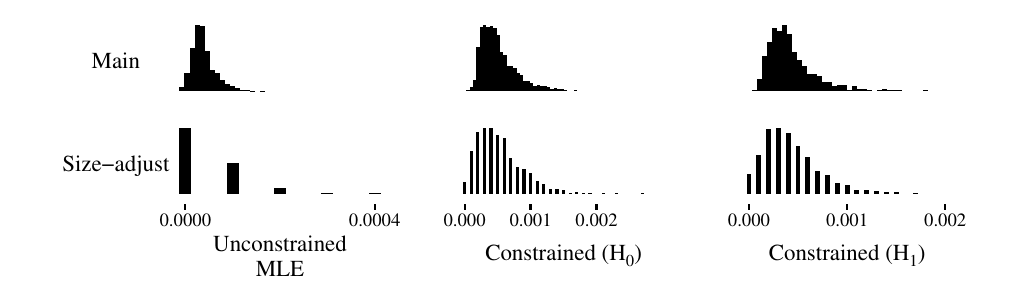}
		\caption{Rates of convergence failures out of one million main samples and ten thousand size-adjust samples at each likelihood fit: respectively, nonconvergence rates in unconstrained MLE, MLE constrained to $H_0: \rho_{AB | C} = \rho_{AB | C}^{\mathrm{null}}$ and MLE constrained to $H_1: \rho_{AB | C} = \rho_{AB | C}^{\mathrm{power}}$.}
		\label{fig:biparcorr-convergence}
		
		\vspace{1em}
		
		\includegraphics[width=\textwidth]{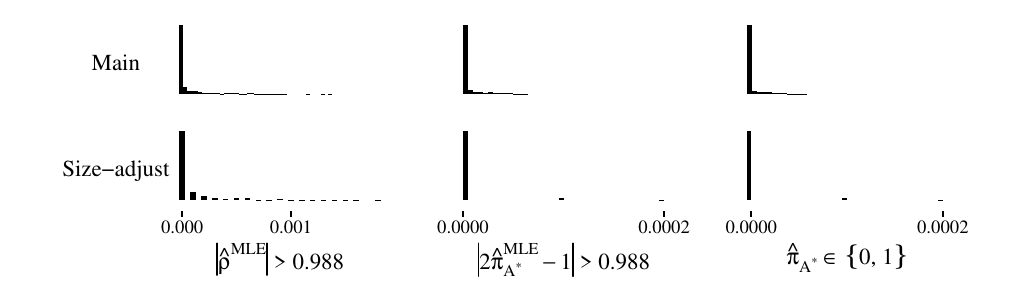}
		\caption{Rates of boundary and degenerate samples out of one million main samples and ten thousand size-adjust samples per parameter sample: respectively, proportion of boundary samples for correlations and prevalences, followed by degenerate samples (those whose $\Astar$ has all the same values).}
		\label{fig:biparcorr-degenerate-and-boundary}
	\end{figure*}

	In Figures \ref{fig:biparcorr-convergence} and \ref{fig:biparcorr-degenerate-and-boundary}, we plot, across the one thousand parameter samples, convergence, boundary and degenerate sample rates, each computed on either one million main or ten thousand size-adjust simulations.  Non-convergence (Figure \ref{fig:biparcorr-convergence}) is broken down by each of the three types of likelihood fitted.  The three boundary and degenerate sample plots in Figure \ref{fig:biparcorr-degenerate-and-boundary} count cases of the following, respectively:
	\begin{enumerate}
		\item \label{rho above 0.99}  At least one correlation MLE lay sufficiently close to the artificially imposed boundary at $|\hat \rho^{\mathrm{MLE}}| = 0.99$ that it may have been distorted.  We define this as any sample with final $|\hat \rho^{\mathrm{MLE}}| > 0.988$.
		
		\item \label{prop_a trans above 0.99}  The prevalence MLE lay sufficiently close to the artificially imposed boundaries at 0.005 and 0.995, that its likelihood may have been distorted.  After accounting for the transform of prevalences onto the range $(-1, 1)$, we define this similarly as any sample with final $|2\hat\pi_{\Astar}^{\mathrm{MLE}} - 1| > 0.988$.
		
		\item All binary $\Astar$ values were of the same value, zero or one.  That is to say, $\hat \pi_{\Astar} \in \{0, 1\}$.  In this case, the biserial correlation is undefined.
	\end{enumerate}

\begin{figure*}[tbp]
		\centering
		\includegraphics[width=\textwidth]{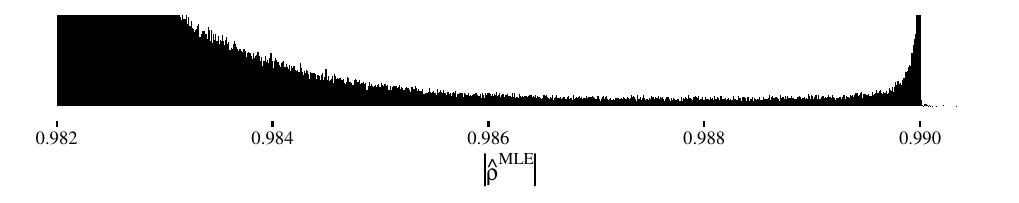}
		\caption{Close-to-boundary absolute values of maximum likelihood estimates pooled across all correlations fitted by maximum likelihood on all runs.  Note that this is very much zoomed in: the $y$-axis here is severely cropped (the maximum count in this plot is 200, vs 25340 if the $y$-limits were left free).}
		\label{fig:biparcorr-extreme-rho}
	\end{figure*}

	The cutoff of 0.988 in step $\ref{rho above 0.99}$ was defined conservatively, based on Figure \ref{fig:biparcorr-extreme-rho}.  Quite visible in this plot is a small build-up of samples just inside the 0.99 boundary.  That is to say, not all boundary points converged precisely to 0.99.  Choosing a threshold of 0.988, then, should be conservative, as this is quite far from the apparent build-up and also includes a number of samples that truly converged above 0.988, without ever testing the boundary.
	
\begin{figure*}[tbp]
		\centering
		\includegraphics[width=\textwidth]{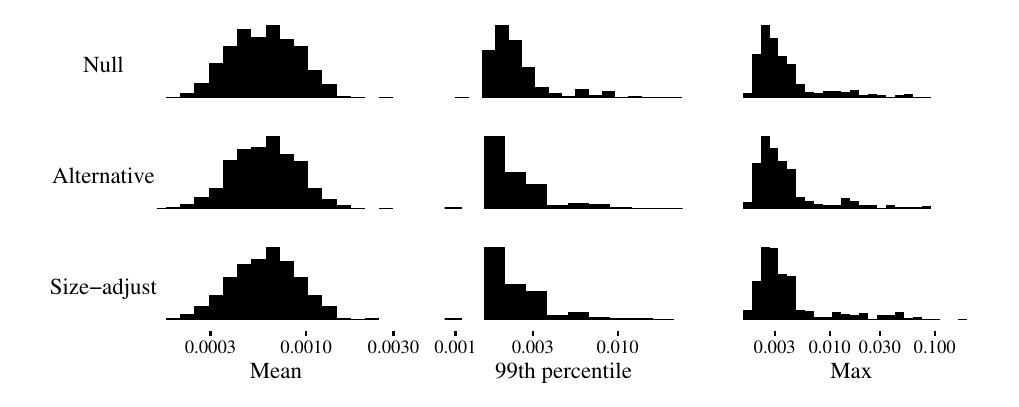}
		\caption{Summaries within each parameter sample across the $p$-values generated at each parameter sample of the non-convergence rates of the bootstrap samples within that $p$-value.  For example, each datapoint on the ``Mean'' histogram is for a single parameter sample, the mean across all $p$-values generated for that parameter sample of the non-convergence rate across all bootstrap samples for the respective $p$-value.}
		\label{fig:bootlr-convergence}
	\end{figure*}

	In Figure \ref{fig:bootlr-convergence}, we summarize the convergence rates across the bootstrap samples generated to convert them into $p$-values in the bootstrap LR.  When bootstrapping, we have 500 parameter samples (index them by $i$), each made up of 500 $p$-values $v$, each in turn made up of 2000 bootstrap profile LR samples $b$.  To summarize these, Figure \ref{fig:bootlr-convergence} plots
	\begin{align*}
		x_i = \mathrm{Summary}_v( \E_b [1 -\mathrm{ converged}_{ivb}])
	\end{align*}
	for $i \in \{1 \ldots 500\}$ and $\mathrm{Summary} \in \{\mathrm{Mean}, \mathrm{99th\ Percentile}, \mathrm{Max}\}$.

	\clearpage
	\FloatBarrier
	\subsection{Deviation from the maximal invariant}
	\label{section sine search further training}

\begin{figure*}[tbp]
		\centering
		\includegraphics[width=\textwidth]{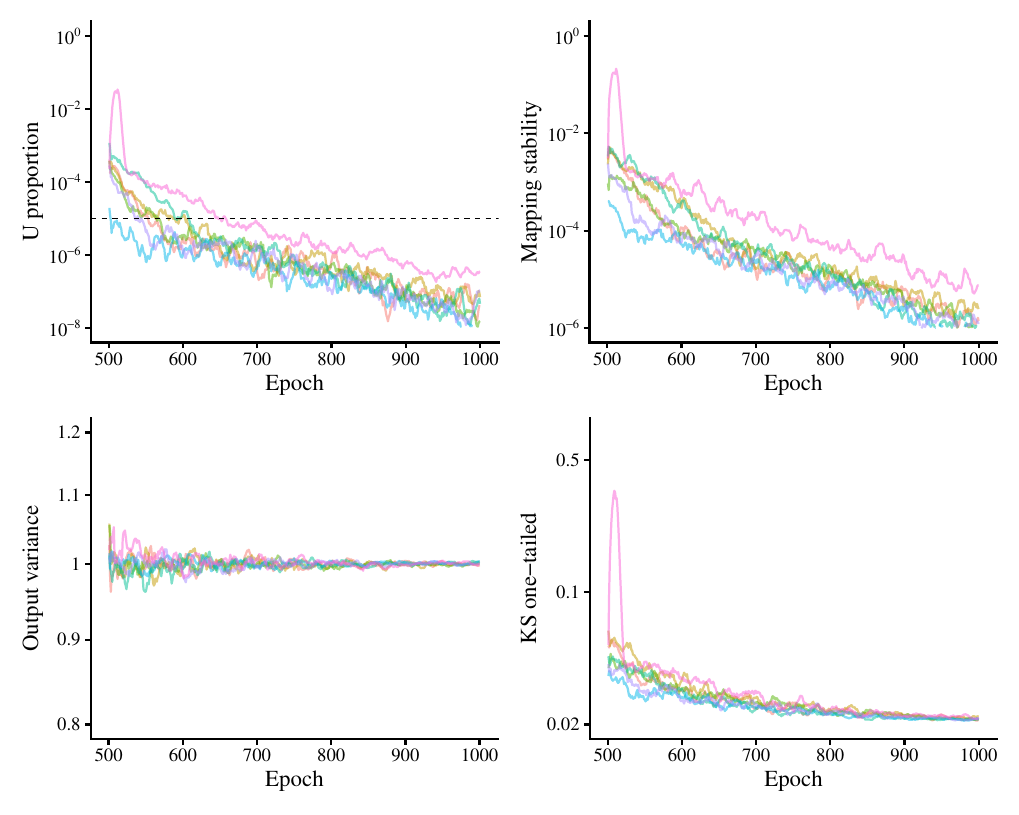}
		\caption{Training the coloured runs from Figure \ref{fig:u-dependence} for a fresh set of 500 epochs of 100 steps.}
		\label{fig:u-dependence-trained-more}
	\end{figure*}

	In Figure \ref{fig:u-dependence-trained-more} we present the effect of further training the seven slower converging runs from Section \ref{section results toy problem} (that is, the coloured curves in Figure \ref{fig:u-dependence}) for a fresh set of 500 epochs of 100 steps.  Once the training runs find their way into the right direction, they converge as neatly as the other 243 runs did in the first set of training epochs.  The further epochs (starting at epoch 500) start again from the initial higher learning rate.
	
\end{document}